\documentclass[useAMS,usenatbib]{mnras}

\newcommand{\lc}{$\ell$\,Carinae}
\newcommand{\kms}{km\,s$^{-1}$}
\newcommand{\ms}{m\,s$^{-1}$}
\newcommand{\drp}{$\Delta R/p$}
\newcommand{\Ppuls}{$P_{\rm{puls}}$}
\newcommand{\Coralie}{{\it Coralie}}

\usepackage{journal_font}
\usepackage{graphicx}
\usepackage{hyperref}
\usepackage{txfonts}
\begin{document}

\title[Modulated Line Variability and Velocity Gradients]{Discovery of Cycle-to-cycle Modulated Spectral Line Variability and
Velocity Gradients in Long-period Cepheids \thanks{Based on observations collected using the
CORALIE echelle spectrograph mounted to the Swiss 1.2m Euler telescope located at ESO La Silla Observatory, Chile.} }

\author[R.I.~Anderson]{Richard I.
Anderson$^{1,2,3}$\thanks{E-mail:\texttt{ria@jhu.edu}}\\
$^1$Physics and Astronomy Department, The Johns Hopkins University, 3400 North
Charles St, Baltimore, MD 21202, USA \\
$^2$Swiss National Science Foundation Fellow\\
$^3$D\'epartement d'Astronomie, Universit\'e de Gen\`eve, 51 Ch. des
Maillettes, CH-1290 Sauverny, Switzerland}
%

\pagerange{\pageref{firstpage}--\pageref{lastpage}} 

\pubyear{2015}

\maketitle 

\label{firstpage}

\maketitle

\begin{abstract}

This work reports the discovery of cycle-to-cycle modulated spectral line and
atmospheric velocity gradient variability in long-period Cepheids based on 
925 high-resolution optical spectra of $\ell$\,Carinae ($P \sim 35.5$\,d) 
recorded during three heavy duty-cycle monitoring campaigns (in 2014,
2015, and 2016). Spectral line variability is investigated via  
cross-correlation functions (CCFs) computed using three sets of spectral lines
(weak, solar, strong). A metallic line velocity gradient, $\delta v_r (t)$, is
computed as the difference between weak and strong-line RVs. CCF shape indicators BIS (asymmetry), FWHM, and depth all exhibit clear
phase-dependent variability patterns that differ from one pulsation cycle to the
next. Weak-line CCFs exhibit these effects more clearly than strong-line CCFs.
BIS exhibits the most peculiar modulated variability and can be used
to identify the presence of cycle-to-cycle modulated line profile variations.
$\delta v_r (t)$ clearly exhibits cycle-to-cycle differences that correlate very
closely with modulated BIS variability, suggesting perturbations of the
atmospheric velocity field as the cause for modulated spectral
line variability. These perturbations are most significant during
contraction and are not in phase with the pulsation, transmitting information
between consecutive pulsation cycles.
This work shows RV curve modulation to be a consequence of atmospheric velocity
gradient perturbations. Possible origins of these perturbations and
their impact on Cepheid RV measurements as well as the projection
factor used in Baade-Wesselink-type distance determinations are discussed.

\end{abstract}


%

\section{Introduction}
Classical Cepheid variable stars (henceforth: Cepheids) are of great interest
for several astrophysical and cosmological applications. This includes
calibrating the extragalactic distance scale with unprecedented accuracy
\citep{2016arXiv160401424R} and serving as high-sensitivity test beds for
state-of-the-art stellar evolution models
\citep[e.g.][]{2016A&A...591A...8A}.
Cepheids provide crucial insights into stellar structure and oscillations
thanks to their high-amplitude radial oscillations that can be studied
photometrically \citep{1786RSPT...76...48G}, spectroscopically
\citep{1894AN....136..281B}, and interferometrically
\citep{2000ApJ...543..972N,2001A&A...367..876K}. The long-period
Cepheid $\ell$\,Carinae is a particularly interesting specimen, since its
variability can be resolved with great precision using all three methods thanks
to its brightness and large angular diameter
\citep[][henceforth: A16]{2016MNRAS.455.4231A}.

Cepheid variability is frequently thought to be well-understood and relatively
simple. While this is true in comparison with other types of stellar
variability, recent advances in instrumentation are revealing exciting new
features of Cepheid pulsations. Of particular relevance for this work is
\emph{modulated variability}, i.e., irregularities of the variability that can
occur between consecutive pulsation cycles\hbox{---}referred to here as
cycle-to-cycle modulation\hbox{---}as well as on longer timescales (months to
years). 

Modulated variability was discovered in the sole classical Cepheid located
inside the original {\it Kepler} field, V1154\,Cygni, whose pulsation period and
amplitude vary rapidly \citep{2012MNRAS.425.1312D}. Two further Cepheids
observed with space-based high-quality photometry\hbox{---}via the {\it MOST}
satellite\hbox{---}were also shown to exhibit cycle-to-cycle changes
\citep{2015MNRAS.446.4008E}. However, detecting such variability is not
straightforward, even with photometry from space 
\citep{2015MNRAS.454..849P}.
Longer temporal baselines, albeit with lower photometric precision, are
achievable from the ground, and data from the {\it Optical Gravitational Lensing
Experiment (OGLE)} have been used to identify peculiarities in the light curves
and frequency spectra of hundreds of Cepheids
\citep{2008AcA....58..163S,2015AcA....65..297S,2015AcA....65..329S}, the
majority of which are short-period Cepheids pulsating in the first overtone.

These new indications of additional complexity in Cepheid pulsations have to be
considered in the context of similar results obtained for other types of
pulsating stars, such as $\delta$\,Sct and $\gamma$\,Dor stars 
\citep[e.g.][]{2016MNRAS.460.1970B,2016arXiv160504443G}.
Additionally, the well-known \citet{1907AN....175..325B} effect among RR\,Lyrae
stars has been studied in exquisite detail using {\it Kepler} photometry
\citep{2010ApJ...713L.198K,2010MNRAS.409.1244S}. 

Recently, \citet[henceforth: A14]{2014A&A...566L..10A} reported the discovery of
modulated radial velocity (RV) variability of four Cepheids, two of
which\hbox{---}QZ\,Nor and V335\,Pup\hbox{---}are short-period Cepheids likely
to pulsate in the first overtone and candidates for Blazhko-effect-like
long-timescale modulations (years) similar to the enigmatic V$473$\,Lyr
\citep{1982A&A...109..258B,2013AN....334..980M}.
The two additional Cepheids presented in A14\hbox{--}$\ell$ Car and
RS\,Pup\hbox{--}are long-period Cepheids for which significant RV curve
modulations among consecutive pulsation cycles were detected, varying in
particular the RV amplitude. For $\ell$\,Car, this effect was
further investigated in a campaign combining contemporaneous
spectroscopy and long-baseline near infrared (NIR) interferometry (A16).

Modulated RV variability represents a significant difficulty for at least two
types of studies involving Cepheids. Firstly, the change in RV amplitude
translates into time-dependent RV curve integrals, which represent a systematic
uncertainty for \citet{1926AN....228..359B}-\citet{1946BAN....10...91W} (BW)
distance determination (A14). Additionally, projection factors required by BW
methods appear to be subject to a complex time (cycle-to-cycle) dependence,
since modulation affects RV variability and angular diameter variations
differently (A16).
Secondly, RV curve modulation acts as noise for the detection of low-mass
companions to Cepheids, at times leading to apparent time variations in the
pulsation-averaged velocity $v_\gamma$.
This complicates the interpretation of time variable $v_\gamma$ on the order of
a few hundred \ms\ and impacts the determination of upper limits for non-binary
Cepheids 
\citep[cf.][and R~.I.~Anderson et al. submitted]{2015AJ....150...13E}.

The origin of modulated variability in Cepheids is currently largely unclear,
although different mechanisms that have been suggested may be related, such as
strange-mode and non-radial pulsations, magnetic cycles, or granulation
\citep{1997A&A...326..669B,2001ApJ...555..961B,2003A&A...401..661K,2009ApJ...696L..37S,2014A&A...563L...4N}.
As mentioned in A14, the very different timescale of RV curve modulation found
in long and short-period Cepheids suggests that multiple mechanisms may be
at play.
Additionally, the link between RV curve modulation and photometric period and
amplitude variations remains as yet to be established. High-precision
photometric observations of $\ell$\,Car using the {\it BRITE} satellites could close this
important gap.

This paper aims at providing new insights into the origin of such complex
cycle-to-cycle modulations by investigating spectral line shape variability. As
a first step towards a more detailed description based on individual lines, it
focuses on cross-correlation functions (CCFs) computed for 925 high-quality,
high-resolution optical spectroscopic data of $\ell$\,Car. This is appropriate,
since CCFs are used to infer high-precision RVs among which RV curve modulation
was discovered. CCFs have the further benefit of greatly enhanced
signal-to-noise ratios (SNR) compared to individual spectral lines and are thus
well-suited to search even for weak signs of modulated variability.

The structure of the paper is as follows. Section\,\ref{sec:obs} describes
the observational data and 
presents the quantities investigated in the following sections, 
including proxies for CCF asymmetry as well as RVs computed at different atmospheric levels
and the impact of different RV measurement methods.
Section\,\ref{sec:results} presents a comprehensive overview of $\ell$\,Car's
modulated spectroscopic variability, starting with RVs based on different
measurement definitions in \S\ref{sec:RVs}. \S\ref{sec:CCFs} then illustrates
line profile variability as a function of phase as well as cycle-to-cycle
modulation using CCF shape indicators BIS, FWHM, and normalized depth, with a
special focus on the asymmetry parameter BIS.
\S\ref{sec:RVgrad} describes in detail the phase variability of the metallic
line velocity gradient as well as its cycle-to-cycle modulation and relation to
BIS. Section\,\ref{sec:disc} discusses possible astrophysical origins of the
discovered behavior (\S\ref{disc:gradients}), consequences for
Cepheid RV measurements (\S\ref{disc:RVmeas}), and implications for BW distance
determination, in particular related to projection factors
(\S\ref{disc:pfactors}). The final section\,\ref{sec:summary} summarizes the
results and concludes.


\section{Observations, velocities \& CCFs}\label{sec:obs}

\subsection{\Coralie\ monitoring campaigns}\label{sec:obs:Coralie}

All data presented here are based on optical spectra of $\ell$\,Carinae observed
with the high-resolution ($R \sim 60\,000$) echelle spectrograph \Coralie\
\citep{2001Msngr.105....1Q,2010A&A...511A..45S}, which is mounted to the
$1.2$\,m Swiss Euler telescope situated at La Silla Observatory, Chile. \Coralie's dedicated data
reduction pipeline performs bias correction, flatfielding, and cosmic ray
removal. The wavelength solution is supplied by a ThAr lamp. \Coralie\ is housed
in a thermally controlled room and any small intra-night variations in
wavelength solution are corrected using reference
spectra recorded simultaneously with the science exposure to reach
single-digit \ms\ RV precision \citep[e.g.][]{2013A&A...551A..90M}.

\Coralie\ has been upgraded twice over the course of the observations (2014 to
2016). Upgrades implemented in late 2014 are described in A16 and 
have had $\sim 15\,$\ms\ impact on RV zero-point,
which is only marginally relevant for the present work that deals with
variations larger by one to three orders of magnitude.
In mid 2015, the method for intra-night wavelength drift
correction was changed.
Formerly, the simultaneous wavelength reference was supplied by a ThAr lamp; now
it is supplied by a Fabry-P\'erot interferometer (FP), which further increased
RV precision.
The nightly wavelength calibration continues to be provided by a ThAr lamp, and
the FP operates relative to this solution. Hence, no significant zero-point
offset is expected, and none has been found thus far. A further
exchange of \Coralie's CCD controller has also had no impact on the RV
zero-point (F. Pepe and Geneva exoplanet group, priv. comm.) while 
reducing read-out time and improving noise properties.

This paper peruses \Coralie\ spectra observed for two prior publications
(A14: 2014 data, A16: 2015 data)
as well as $230$ new, as yet unpublished observations carried out between 2016
February 12 and 2016 May 10. Each of the three epochs (2014, 2015, 2016) considered here
span two complete consecutive pulsation cycles.
This allows to investigate cycle-to-cycle behavior as well as long-term changes
in the RV and spectral variability.

\subsection{Cross-correlations and inferred RV}\label{sec:obs:CCFs+RVs}

All RVs presented here are measured using the cross-correlation
technique \citep{1996A&AS..119..373B,2002A&A...388..632P} and are expressed
relative to the solar system barycenter.
This technique cross-correlates a weighted numerical mask and the observed
spectrum to produce cross-correlation functions (CCFs) based on which RV is
measured by fitting Gaussian profiles. The line mask used for most Cepheid
observations is representative of a solar spectral type (henceforth: G2 mask),
and CCFs are computed such as to resemble absorption lines
\citep[for details, see][]{2002A&A...388..632P}. These velocities are referred
to as ``Gaussian RVs'' with symbol $v_r$. 

\begin{figure}
\centering
\includegraphics{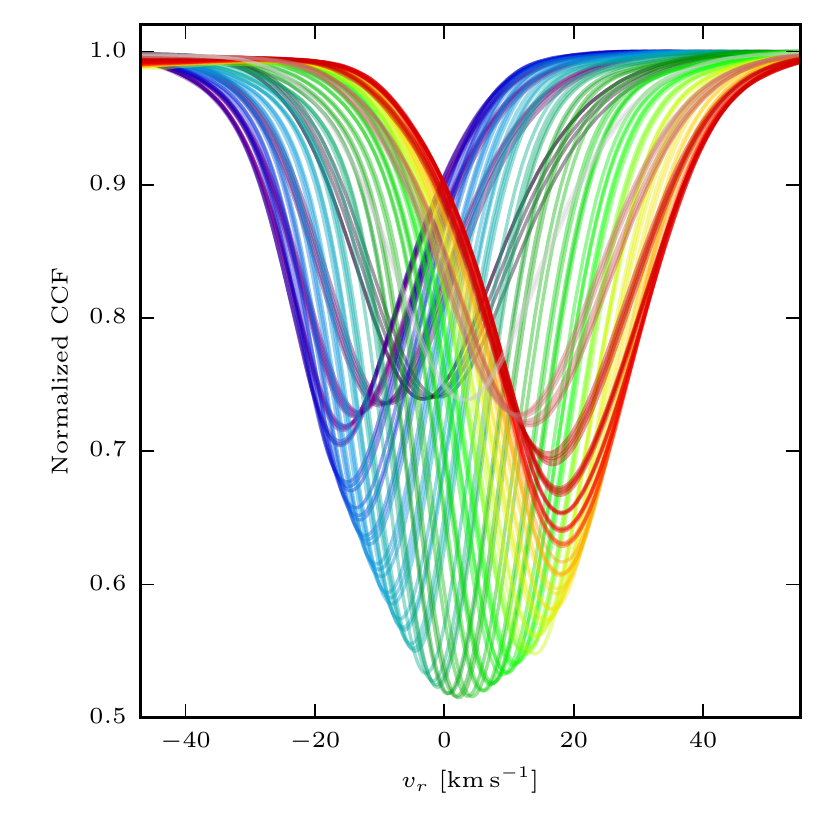}
\caption{Cross-correlation functions (CCFs) calculated using a G2 line mask and
data from the first complete 2016 cycle. Colors trace pulsation phase,
the cycle proceeds from the top center counterclockwise.}
\label{fig:CCFvariability}
\end{figure}

Figure\,\ref{fig:CCFvariability} shows CCFs computed using the first fully
sampled pulsation cycle of the 2016 campaign. The pulsation cycle proceeds
counterclockwise and pulsation phase is color-coded to aid visual inspection.
Fig.\,\ref{fig:CCFvariability} clearly shows several phase-dependent features,
including changes in: a) temperature (spectral type) via the varying depth of
the CCF; b) RV associated with the pulsation via the displacement of the CCF
along the abscissa; c) the width of the CCF; d) CCF asymmetry.

As a consequence of line asymmetry, RVs measured by fitting profiles such as
Gaussian functions to CCFs (as well as to individual spectral lines) are
biased \citep{1982A&A...109..258B} and much work has been done with the
aim of improving the accuracy of Cepheid RV measurements by considering line
asymmetry or different velocity curves of different spectral lines
\citep[e.g.][]{1967IAUS...28..207K,1990ApJ...362..333S,1992MNRAS.259..474W,1993ApJ...415..323B,2000MNRAS.314..420K,2006A&A...453..309N,2007PASP..119..398G}.
Nevertheless, Gaussian RVs are extremely precise in that they are able to
reproduce a consistent value under identical conditions, and this is illustrated
by very smooth variations and small scatter when investigated appropriately (cf.
A16). It is thanks to this precision that RV observations are now revealing
previously unknown complexity in Cepheid pulsations.
 
This level of extreme RV precision should however not be confused with accuracy,
i.e., the ability to reproduce the ``true'' value precisely. This is primarily
because Cepheid atmospheres are highly dynamical and thus not characterized by a
single velocity at a given phase. CCF-based RVs represent a weighted average RV
of thousands of lines formed at different levels in the atmosphere and are
therefore difficult to interpret in detail.
However, even individual spectral lines are not free of such difficulty, since
line formation in supergiant atmospheres occurs over significantly extended
regions and is therefore more susceptible to the velocity field than in dwarf
stars.

\begin{figure}
\centering
\includegraphics{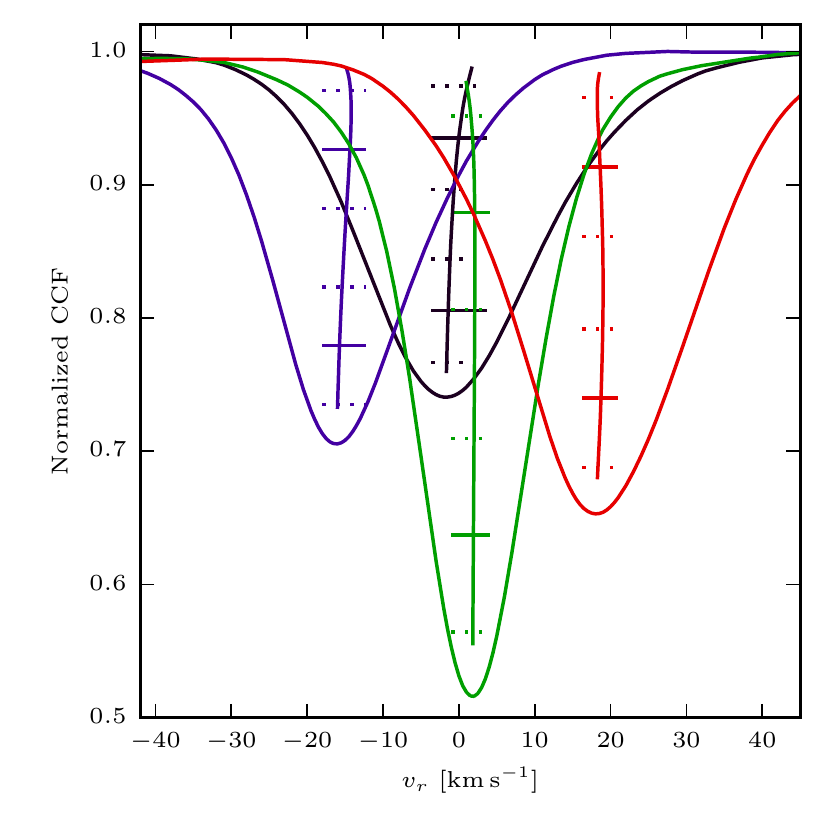}
\caption{Illustration of the quantity BIS for four pulsation phases. BIS is the
velocity difference (on the abscissa) between the CCF bisector (solid nearly
vertical line) near its top and bottom, computed as the average of the regions
included between horizontal dotted lines \citep{2001A&A...379..279Q}.}
\label{fig:BISdefinition}
\end{figure}

Despite these shortcomings, CCFs as well as RV curves based on them contain a
great deal of useful information. In particular, CCF
shape parameters exhibit smooth variations with phase due to the high SNR of
CCFs and allow to investigate the relation between RV curve
modulation and line profile variability.
Specifically, this work considers\hbox{---}in addition to RVs\hbox{---}the
CCFs' full width at half maximum (FWHM) and normalized depth, both of
which are measured by proxy via the fitted Gaussian profile.
In addition, the quantity BIS (bisector inverse span) computed directly via CCFs
serves as a proxy for line asymmetry.
BIS is computed as the difference between average bisector velocities 
near the top $10\hbox{--}40$ and bottom $60\hbox{--}90\,$\% of the CCF, see
Fig.\,\ref{fig:BISdefinition} and \citet[Fig.\,5]{2001A&A...379..279Q}. 

\begin{figure}
\centering
\includegraphics{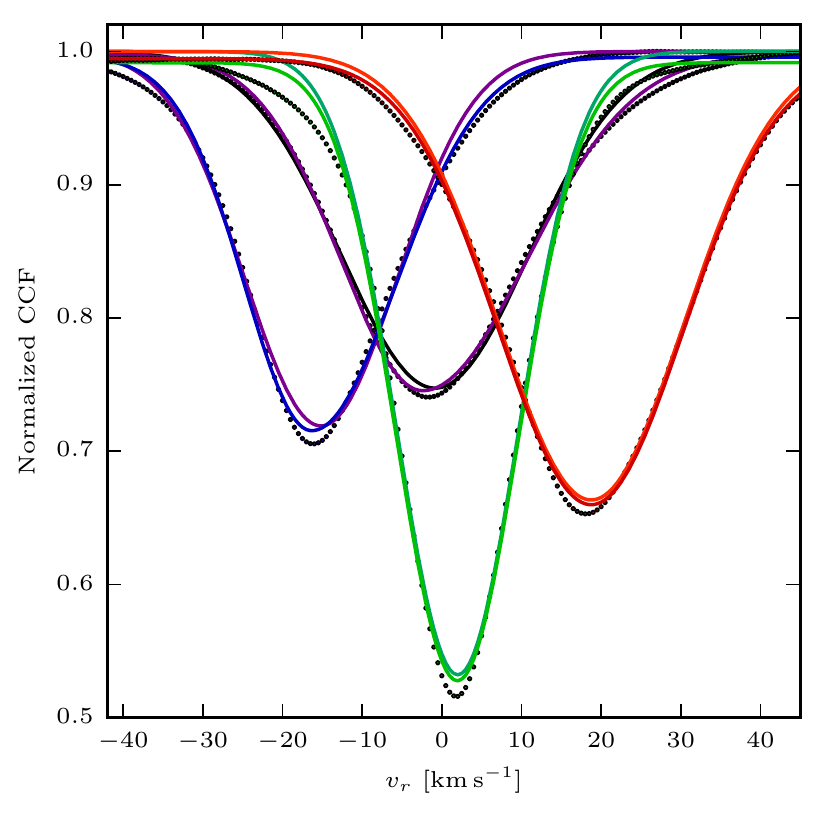}
\caption{Illustration of Gaussian and bi-Gaussian RV profile fitting to CCFs of
four different pulsation phases (selected as minimum radius, maximum expansion
velocity, maximum contrast, and maximum contraction velocity). RV is defined as
negative for objects approaching the observer.}
\label{fig:CCFgaussBigauss}
\end{figure}

Bi-Gaussian profiles have been proposed as an alternative to Gaussian profiles
for measuring RV based on CCFs, specifically to take into account line
asymmetry \citep{2006A&A...453..309N}.
To investigate how this different method of measuring RV on a given CCF reacts
to cycle-to-cycle changes in the spectral line variability, 
bi-Gaussian RVs are determined for all CCFs employed in this work using
the {\tt python}
implementation\footnote{\url{http://bitbucket.org/pedrofigueira/line-profile-indicators}}
by \citet{2013A&A...557A..93F}. Bi-Gaussian RVs are denoted by the symbol
$v_{r,\rm{biG}}$.

Figure\,\ref{fig:CCFgaussBigauss} exemplifies the difference between a Gaussian
and bi-Gaussian fit to four selected CCFs. Close inspection shows 
that bi-Gaussians do trace the computed CCF points more closely than a
Gaussian fit. There does remain, however, a noticeable difference between the
observed and fitted profiles. As a consequence of their tracing the CCF core more
closely than Gaussians, bi-Gaussian RVs can be expected to lead to larger RV
amplitudes.

\subsection{Metallic line velocity gradient}\label{sec:obs:RVgrad}

\begin{table*}
\centering
\caption{Explanation of velocity difference in terms of compression and stretch
within the atmosphere. Upper layers are traced by stronger (deeper) absorption
lines, lower layers by weaker (shallower) lines. $\delta v_r =
v_{r,\rm{strong}} - v_{r,\rm{weak}} - 0.641\,$\kms\ (offset corrects for
differential bias in $v_\gamma$ for weak and strong line RVs).}
\begin{tabular}{l|c|c}
Pulsational motion & expansion ($v_r \lesssim v_\gamma$) & contraction ($v_r
\gtrsim v_\gamma$) \\
Velocity gradient &  &  \\
\hline
$\delta v_r < 0$ & upper layers expand faster:
{\bf stretch} & upper layers contract more slowly: {\bf stretch} \\
$\delta v_r > 0$ & lower layers expand faster: {\bf compression} & lower layers
contract more slowly: {\bf compression} \\
\hline
\end{tabular}
\label{tab:DTexplained}
\end{table*}

It has been long known that Cepheid atmospheres are subject to significant
velocity gradients
\citep[e.g.][]{1956ApJ...123..201S,1967IAUS...28..207K,1969MNRAS.145..377D,1973ApJ...180..895K,1978ApJ...222..578K}
and velocity differences among individual lines have been investigated in detail
\citep[e.g.][]{1992MNRAS.259..474W,1993ApJ...415..323B,2007A&A...471..661N}. 
In this work, a metallic line velocity gradient is computed using two newly-created
correlation masks containing exclusively strong (depth $> 0.65$) and weak (depth
$< 0.55$) lines, respectively. This procedure aims to exploit the benefit of the
superior SNR of CCFs compared to individual lines in order to be maximally
sensitive to cycle-to-cycle modulation.
Both masks are based on the nominal solar (G2 spectral type) mask. The
specific division between weak and strong line masks was 
adopted to achieve a similar weighting of the computed CCFs, i.e., $\sum_{i,\rm{strong}}{ d_i } \sim \sum_{i,\rm{weak}}{ d_i } $, where $d_i$ denotes the line strengths as specified
in the G2 mask. The weak-line mask thus contains 2030 lines, compared to 1209
lines in the strong-line mask. 

Using strong and weak-line RVs measured by fitting Gaussian profiles to CCFs
computed using the strong and weak-line masks, the metallic velocity
gradient is defined as
\begin{equation}
\delta v_r (t) = v_{r,\rm{strong}}(t) - v_{r,\rm{weak}}(t) - 0.641\
\rm{km\,s^{-1}}\ ,
\label{eq:rvgradient}
\end{equation}
with an offset of $0.641\,$\kms\ to correct for  
differential bias in the pulsation averaged velocities, cf. \S\ref{sec:RVs} and
\S\ref{sec:CCFs} as well as the well-known $k-$term problem
\citep[e.g.][]{2008A&A...489.1255N}.
Uncertainties on $\delta v_r (t)$ are computed as the squared sum of each mask's
RV uncertainties.

$\delta v_r (t)$ defined in Eq\,\ref{eq:rvgradient} traces a velocity difference
among lines formed at higher (stronger lines) and lower (weaker lines) levels in the Cepheid
atmosphere \citep[cf.][]{1994A&A...285.1012G}. $\delta v_r (t)$ thus indicates
whether the region over which the gradient is valid is being compressed
(positive $\delta v_r$) or stretched (negative $\delta v_r$) by the
pulsation (as usual, RV is positive when receding from the observer).
Table\,\ref{tab:DTexplained} succinctly summarizes this.
Similar techniques have been employed for Mira stars to investigate shock
propagation \citep{2001A&A...379..288A}. 

\subsection{Variability with phase and modulated variability}
\label{sec:obs:definitions}

This paper describes variations on different timescales, i.e., 1)
variability over a pulsation cycle (alternatively: with pulsation phase, $P \sim
35.5$\,d) and 2) modulated variability, which denotes changes in the former
variability pattern that occur over timescales longer than one pulsation cycle,
ranging from one cycle to the next up to 2 years (baseline of the 
observations).

This work discusses differences between measurements on different timescales,
i.e., data recorded 1) at the same time
or 2) at the same phase during different pulsation cycles.
The following notation is adopted to clearly distinguish these cases.

Differences of quantities observed \emph{at the same time} are labeled as (lowercase) $\delta$.
For instance, the metallic line velocity gradient $\delta v_r (t)$ is the
velocity difference of two different atmospheric layers measured using the same observed
spectrum (cf. \S\ref{sec:obs:RVgrad}). Differences of quantities observed in
\emph{different pulsation cycles} are labeled using (uppercase) $\Delta$. For
instance, the difference in RV curve between two pulsation cycles is $\Delta v_r
(\phi) = v_{r,\rm{cycle2}} (\phi) - v_{r,\rm{cycle1}} (\phi)$. In these cases,
phase $\phi$ is computed using ephemerides determined in \S\ref{sec:timings}
below.

\section{Results}\label{sec:results}

Figure\,\ref{fig:RVcurve} shows the Gaussian RVs measured during the three
monitoring campaigns and illustrates the nomenclature adopted for each
pulsation cycle as used in the following.
Table\,\ref{tab:RVs:data} lists velocities and CCF shape parameters (cf.
\S\ref{sec:obs:CCFs+RVs}) for a subset of the observations; the complete data
table is made publicly available online via the
CDS\footnote{\url{http://cds.u-strasbg.fr/}}. As mentioned in
\S\ref{sec:obs:CCFs+RVs} $v_r$ denotes RVs measured via Gaussian fits 
to CCFs computed using the G2 mask. Other RV definitions are clearly
identified via their subscripts.

\begin{figure*}
\centering
\includegraphics{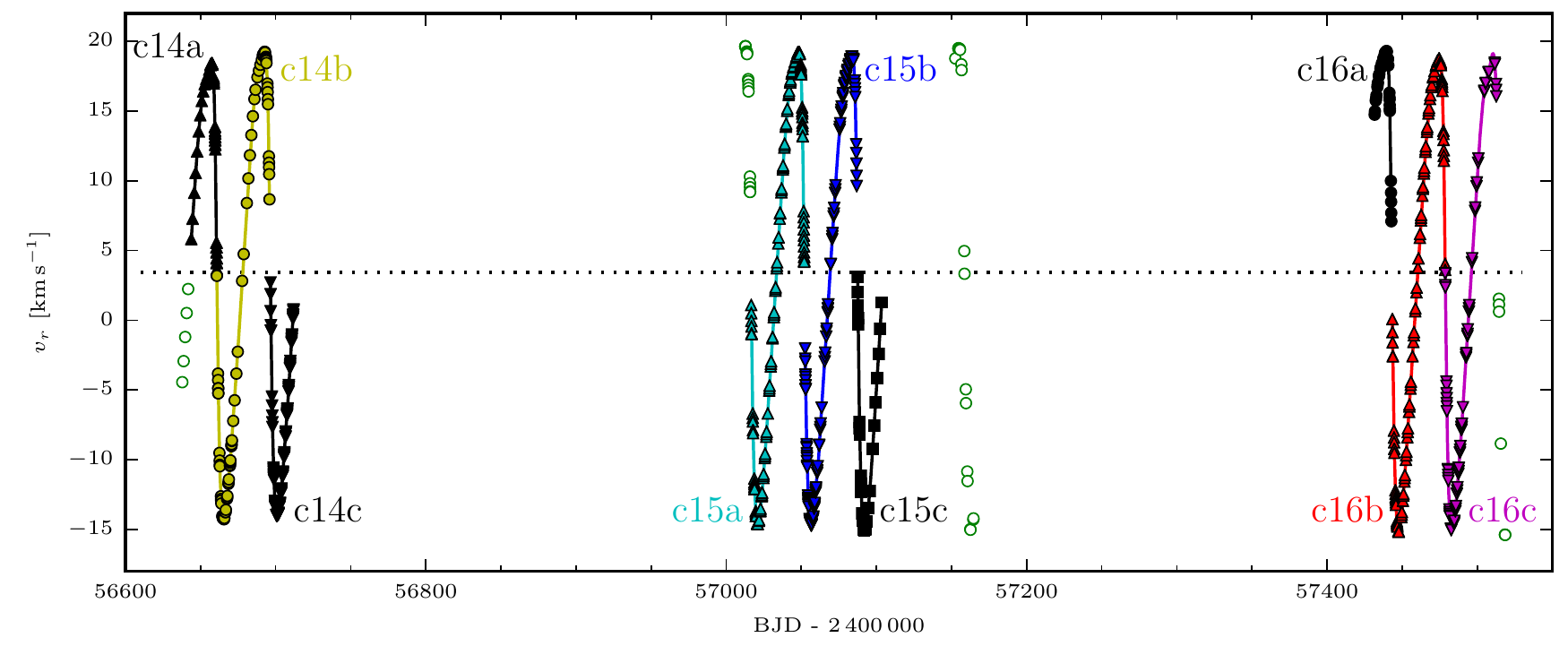}
\caption{RV data of \lc\ from the 2014, 2015, and 2016 campaigns. Individual
pulsation cycles discussed in the following are marked here for comparison with
all other figures. Green open circles show additional data not presented in
detail in the following; these are also made publicly available. The horizontal
dotted line represents $v_\gamma = 3.419$\,\kms\ adopted to determine the
duration of pulsation cycles.}
\label{fig:RVcurve}
\end{figure*}

\begin{table*}
\centering
\begin{tabular}{lrrrrrrrrr}
\hline
BJD - $2\,400\,000$ & FWHM & BIS & depth & $v_r$ &
$v_{r,\rm{biG}}$ & $v_{r,\rm{weak}}$ & BIS$_{\rm{weak}}$ &
$v_{r,\rm{strong}}$ & BIS$_{\rm{strong}}$ \\
 days & [\kms ]  & [\kms ]  & [$\%$] & [\kms ]  & [\kms ] & [\kms ]  & [\kms ] &
 [\kms] & [\kms] \\
\hline
$ 56636.757136 $ & $ 17.4 $ & $ 0.384 $ & $ 41.7 $ & $ -6.233  \pm  0.015 $ & $ -6.578 $ & $ -6.223  \pm  0.015 $ & $ 0.462 $ & $ -6.397  \pm  0.015 $ & $ 0.136 $ \\ 
$ 56637.848761 $ & $ 17.1 $ & $ 0.344 $ & $ 43.2 $ & $ -4.445  \pm  0.015 $ & $ -4.758 $ & $ -4.438  \pm  0.015 $ & $ 0.405 $ & $ -4.603  \pm  0.015 $ & $ 0.084 $ \\ 
$ 56638.748377 $ & $ 17.0 $ & $ 0.318 $ & $ 44.3 $ & $ -2.936  \pm  0.015 $ & $ -3.223 $ & $ -2.947  \pm  0.015 $ & $ 0.375 $ & $ -3.073  \pm  0.015 $ & $ 0.048 $ \\ 
$ 56639.774722 $ & $ 17.0 $ & $ 0.285 $ & $ 45.1 $ & $ -1.205  \pm  0.015 $ & $ -1.463 $ & $ -1.241  \pm  0.015 $ & $ 0.346 $ & $ -1.323  \pm  0.015 $ & $ -0.007 $ \\ 
$ 56640.769897 $ & $ 17.1 $ & $ 0.246 $ & $ 45.5 $ & $ 0.513  \pm  0.015 $ & $0.286 $ & $ 0.445  \pm  0.015 $ & $ 0.352 $ & $ 0.422  \pm  0.015 $ & $ -0.070 $ \\
$ 56641.766439 $ & $ 17.3 $ & $ 0.184 $ & $ 45.7 $ & $ 2.234  \pm  0.015 $ & $2.061 $ & $ 2.120  \pm  0.015 $ & $ 0.310 $ & $ 2.184  \pm  0.015 $ & $ -0.159 $ \\
$ 56643.842825 $ & $ 18.3 $ & $ -0.003 $ & $ 45.1 $ & $ 5.809  \pm  0.015 $ & $ 5.814 $ & $ 5.572  \pm  0.015 $ & $ 0.071 $ & $ 5.873  \pm  0.015 $ & $ -0.342 $ \\ 
$ 56644.719386 $ & $ 18.8 $ & $ -0.135 $ & $ 44.5 $ & $ 7.281  \pm  0.015 $ & $7.41 $ & $ 6.993  \pm  0.015 $ & $ -0.130 $ & $ 7.406  \pm  0.015 $ & $ -0.429 $ \\
$ 56645.850598 $ & $ 19.6 $ & $ -0.316 $ & $ 43.5 $ & $ 9.134  \pm  0.015 $ & $9.431 $ & $ 8.756  \pm  0.015 $ & $ -0.422 $ & $ 9.336  \pm  0.015 $ & $ -0.563 $ \\ 
$ 56646.743305 $ & $ 20.4 $ & $ -0.487 $ & $ 42.6 $ & $ 10.561  \pm  0.015 $ & $11.024 $ & $ 10.109  \pm  0.015 $ & $ -0.713 $ & $ 10.821  \pm  0.015 $ & $ -0.710 $ \\
\multicolumn{10}{c}{\ldots}  \\
$ 57512.458865 $ & $ 28.7 $ & $ -0.476 $ & $ 29.0 $ & $ 16.941  \pm  0.003 $ & $ 17.393 $ & $ 15.237  \pm  0.007 $ & $ -0.316 $ & $ 17.674  \pm  0.004 $ & $ 1.196 $ \\ 
$ 57512.591455 $ & $ 28.8 $ & $ -0.506 $ & $ 28.7 $ & $ 16.514  \pm  0.005 $ & $
17.001 $ & $ 14.755  \pm  0.010 $ & $ -0.340 $ & $ 17.269  \pm  0.005 $ & $ 1.200$ \\
$ 57512.708641 $ & $ 28.9 $ & $ -0.543 $ & $ 28.4 $ & $ 16.057  \pm  0.004 $ & $ 16.632 $ & $ 14.328  \pm  0.009 $ & $ -0.362 $ & $ 16.819  \pm  0.005 $ & $ 1.289 $ \\ 
$ 57514.460114 $ & $ 28.2 $ & $ 1.159 $ & $ 25.1 $ & $ 1.539  \pm  0.003 $ & $0.251 $ & $ 0.259  \pm  0.006 $ & $ -0.350 $ & $ 2.004  \pm  0.003 $ & $ 1.285 $ \\
$ 57514.498343 $ & $ 28.1 $ & $ 1.169 $ & $ 25.2 $ & $ 1.125  \pm  0.005 $ & $ -0.192 $ & $ -0.111  \pm  0.012 $ & $ -0.318 $ & $ 1.567  \pm  0.006 $ & $ 1.375 $ \\ 
$ 57514.547207 $ & $ 28.0 $ & $ 1.208 $ & $ 25.2 $ & $ 0.614  \pm  0.010 $ & $
-0.729 $ & $ -0.530  \pm  0.022 $ & $ -0.324 $ & $ 1.019  \pm  0.012 $ & $ 1.380$ \\
$ 57515.718169 $ & $ 25.9 $ & $ 1.185 $ & $ 25.2 $ & $ -8.849  \pm  0.004 $ & $ -9.984 $ & $ -9.247  \pm  0.008 $ & $ -0.306 $ & $ -8.932  \pm  0.004 $ & $ 1.421 $ \\ 
$ 57518.468137 $ & $ 23.3 $ & $ 1.246 $ & $ 27.4 $ & $ -15.377  \pm  0.004 $ & $ -16.564 $ & $ -15.492  \pm  0.007 $ & $ -0.289 $ & $ -15.597  \pm  0.004 $ & $ 1.436 $ \\ 
$ 57518.573887 $ & $ 23.2 $ & $ 1.229 $ & $ 27.5 $ & $ -15.392  \pm  0.003 $ & $ -16.591 $ & $ -15.499  \pm  0.006 $ & $ -0.297 $ & $ -15.616  \pm  0.003 $ & $ 1.509 $ \\ 
$ 57518.580913 $ & $ 23.2 $ & $ 1.233 $ & $ 27.5 $ & $ -15.396  \pm  0.003 $ & $ -16.611 $ & $ -15.515  \pm  0.006 $ & $ -0.299 $ & $ -15.617  \pm  0.004 $ & $ 1.485 $ \\ 
\hline
\end{tabular}
\caption{Example of the \Coralie\ RV data used here. These data are based on
observations taken in 2014 (A14), and 2015 
(A16), as well as new data from a 2016 campaign.
Measurements for the first and last 10 observations are shown. The full
data set is made publicly available through the CDS. BJD denotes
barycentric Julian date. Columns FWHM and depth are based on the Gaussian
profile fitted to the CCF. BIS denotes bisector inverse span and is measured on
the CCF.
$v_r$ is the RV measured via a Gaussian fit to the CCF computed using
the G2 mask.
$v_{r,\rm{biG}}$ is analogously measured via a bi-Gaussian profile
\citep{2013A&A...557A..93F}.
$v_{r,\rm{weak}}$ and $v_{r,\rm{strong}}$ denote RVs measured by fitting
Gaussians to CCFs computed using weak- and strong-line correlation masks,
respectively. }
\label{tab:RVs:data}
\end{table*}

\subsection{Cycle timing and modulated RV variability}

\subsubsection{Cycle timing}
\label{sec:timings}

$\ell$\,Carinae's pulsation period fluctuates from one pulsation cycle to the
next (e.g. A14, A16). Pulsation ephemerides are thus most precisely
determined using the RV data of each individual pulsation cycle.

This work defines the start of a pulsation cycle to occur at minimum radius,
since the steep RV variation during this phase allows for the most precise
timing measurement \citep[cf.][]{2012MNRAS.425.1312D,2016MNRAS.455.4231A}.
Minimum radius, by definition, is reached when $v_r = v_\gamma$ while $v_r$ is
decreasing, with $v_\gamma$ denoting the pulsation-averaged velocity. The main
uncertainty related to timing the pulsation is therefore the ability to
precisely define $v_\gamma$, since $v_\gamma$ can exhibit erratic temporal
variations due to the effects considered in this paper (see also A16).
Furthermore, $v_\gamma$ depends on the definition of RV employed (cf.
\S\ref{sec:obs:CCFs+RVs}), since $v_\gamma$ is biased due to line asymmetry and
different lines or measurement techniques differ in sensitivity to this bias.
The most consistent way of timing the pulsation via RVs is therefore to
determine $v_\gamma$ separately for each type of RV definition.
Fourier series fits with 13 harmonics to all available \Coralie\ data thus yield
$v_\gamma = 3.419\,$\kms\ for Gaussian and $v_{\gamma,\rm{biG}}=3.441$\,\kms\
for bi-Gaussian RVs based on the G2 mask, as well as $v_{\gamma,\rm{strong}} =
3.571$\,\kms\ and $v_{\gamma,\rm{weak}} = 2.930$\,\kms. For the purpose of
timing the pulsations, these are adopted as \emph{true} values, although the
(statistical) uncertainty of each of these pulsation-averaged velocities is on
the order of $0.05$\,\kms. This comparison also illustrates the systematic
difficulty of determining the absolute systemic velocity to better than a few
hundred \ms.

Specifically, the duration of a pulsation cycle is determined as the time span
between consecutive intersections of the spline-interpolated RV curve with
$v_\gamma$ at minimum radius.
For the 2015 pulsation cycles, the timing specified in A16 is adopted.
Table\,\ref{tab:timings} specifies all cycle timings determined using
Gaussian RVs.

Some of the available spectra were observed outside the date range of
fully traced pulsation cycles. To make use of these data,  
\emph{half-cycles} c14a, c14c, c15c, and c16a are defined as either beginning or
ending at minimum radius. Since the duration of such half-cycles cannot be
determined, a fixed pulsation period of $35.5\,$d is adopted to compute
the corresponding pulsation phase.
Throughout the paper, each cycle is plotted using a consistent scheme of colors
and symbols as shown in Fig.\,\ref{fig:RVcurve}. Half-cycles are drawn in black.

\begin{table}
\centering
\begin{tabular}{lrrr}
Cycle & BJD begin & BJD end & Duration \\
 & $-2\,400\,000$ & $-2\,400\,000$ & [d]  \\
\hline
c14a$^\dagger$ & 56642.448 & 56660.832 & \hbox{---} \\
c14b & 56660.832 & 56696.451 & 35.619 \\
c14c$^\dagger$ &  56696.451 & 56713.398 & \hbox{---} \\
\hline
c15a & 57016.386 & 57051.957  & 35.571 \\
c15b & 57051.957 & 57087.491  & 35.534 \\
c15c$^\dagger$ & 57087.491 & 57104.522 & \hbox{---} \\
\hline
c16a$^\dagger$ & 57431.698 &  57443.184 & \hbox{---} \\
c16b & 57443.184 & 57478.681  & 35.497 \\
c16c & 57478.681 & 57514.278 & 35.597 \\
\hline
\end{tabular}
\caption{Timings of pulsation cycles via  Gaussian RVs based on CCFs computed
using the G2 mask.
Phase is defined such that $\phi = 0$ at minimum radius. $^\dagger$ marks
incompletely traced pulsation cycles. Timings for 2015 data (c15a through c15c)
are adopted from A16.} 
\label{tab:timings}
\end{table}

\subsubsection{Modulated RV variability}\label{sec:RVs}

\begin{figure*}
\centering
\includegraphics{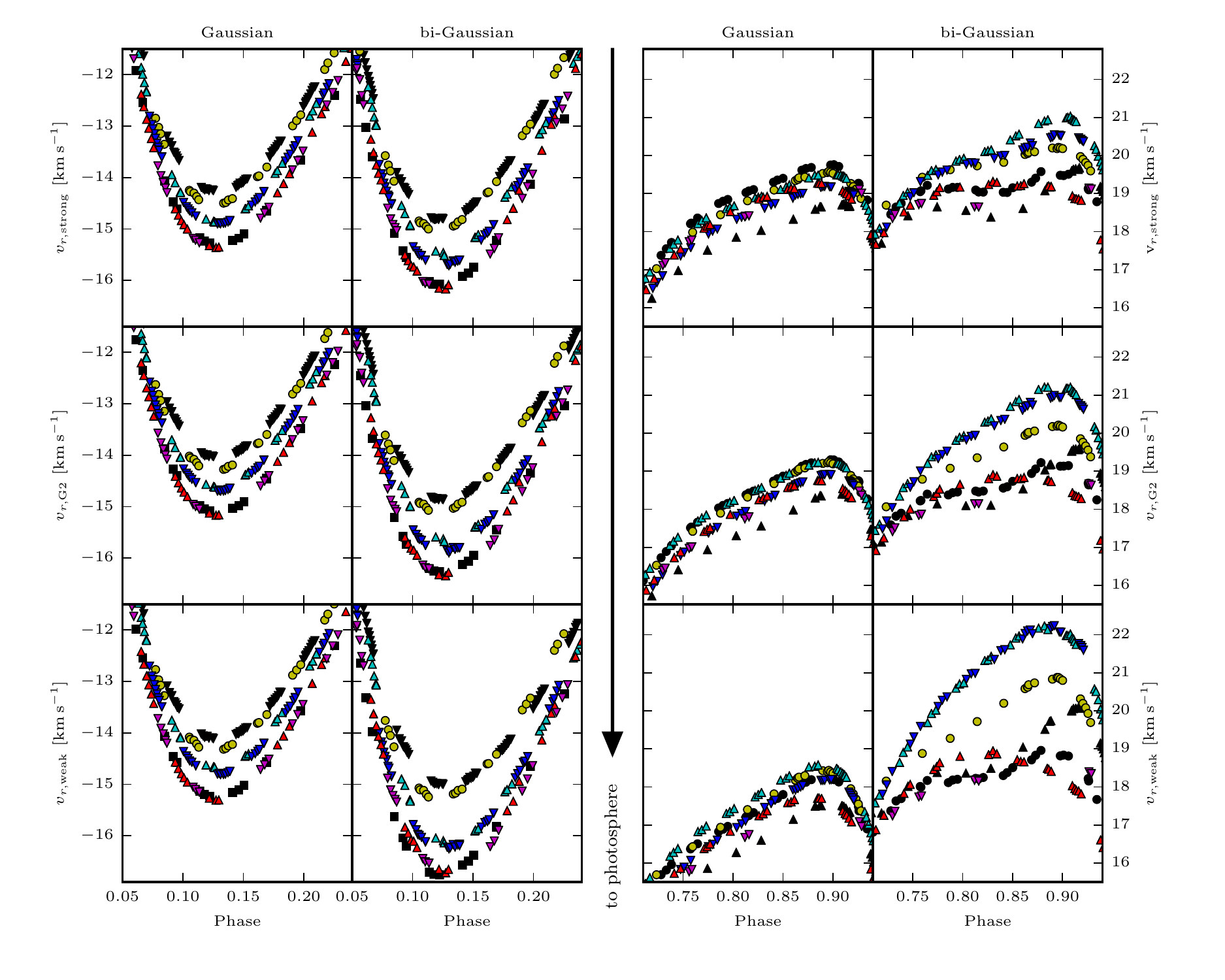}
\caption{Modulated RV variability near minimum (left-hand side) and maximum
(right-hand side) velocity based on lines formed at different heights in the
atmosphere as measured using Gaussian and bi-Gaussian profiles.
RVs shown are based on strong lines in {\it top panels}, the default G2 mask in
{\it center panels}, and weak lines in {\it bottom panels}. 
Symbols distinguish between the pulsation cycles, cf. Fig.\,\ref{fig:RVcurve}.
Statistical errors are smaller than symbol size.}
\label{fig:RVmodulationZoomin}
\end{figure*}

Figure\,\ref{fig:RVmodulationZoomin} illustrates $\ell$\,Car's RV curve
modulation in detail. It shows both Gaussian and bi-Gaussian RVs based on G2,
weak-line, and strong-line correlation masks. The figure shows only the enlarged
sections of the RV curve near minimum and maximum RV, since the modulated RV
variability shows most clearly at these phases.
Fig.\,\ref{fig:RVmodulationZoomin} clearly demonstrates that RV curve modulation
is exhibited regardless of the measurement method or line mask used, although
its amplitude (or extent) depends on the definition of both.

Figure\,\ref{fig:RVmodulationZoomin} also shows that cycle-to-cycle differences
are seen in each of the three campaigns. However, Gaussian RVs from
2015 and 2016 reveal the tendency of longer-timescale modulations to be
stronger than cycle-to-cycle modulations. 2014 data
constitute an exception by differing particularly strongly 
between c14a and c14b near maximum RV.
At minimum RV, the greatest difference seen is between c14c and c15c, followed
by c14b and either of c16b or c. Near maximal RV, the overall greatest
difference is seen for c14a and c16a, whereas the differences among all
other cycles are much weaker.

Bi-Gaussian RVs mirror these variations closely during expansion (near minimum
RV), albeit with greater amplitude. During contraction however, bi-Gaussian RVs
exhibit different RV curve shapes with greatly amplified cycle-to-cycle
differences.
This behavior further strongly depends on the types of lines included in the
CCF, with weak lines leading to much stronger cycle-to-cycle differences than
strong lines due to enhanced asymmetry (\S\ref{sec:CCFs}). 

\subsection{CCF Variations}\label{sec:CCFs}

CCFs are frequently considered as representations of an average spectral line
profile. However, this represents a crude approximation and does not account for
the weighting of the spectral lines applied via the line masks, even for
non-pulsating stars.
In Cepheids, individual spectral lines are known to exhibit significant
phase-dependent asymmetry. Moreover, these asymmetric lines formed at different
depths are moving at different velocities due to the phase-dependent
velocity field. All such lines are summed into a common CCF profile, whose
detailed physical interpretation is thus complicated. 

Nevertheless, certain features of the pulsations, such as temperature
variations, are clearly evident in CCFs, see Fig.\,\ref{fig:CCFvariability}. Hence, CCFs do
remain useful to investigate the variability of line profiles in Cepheids,
although one must be careful to avoid over-interpretation of these
variations.
Of course, CCFs  have the added benefit of very high SNRs, allowing to compare
line profile variability even when the SNR per pixel of the spectra is rather
low (down to spectral SNR of $\sim 10$).

\begin{figure*}
\centering
\includegraphics{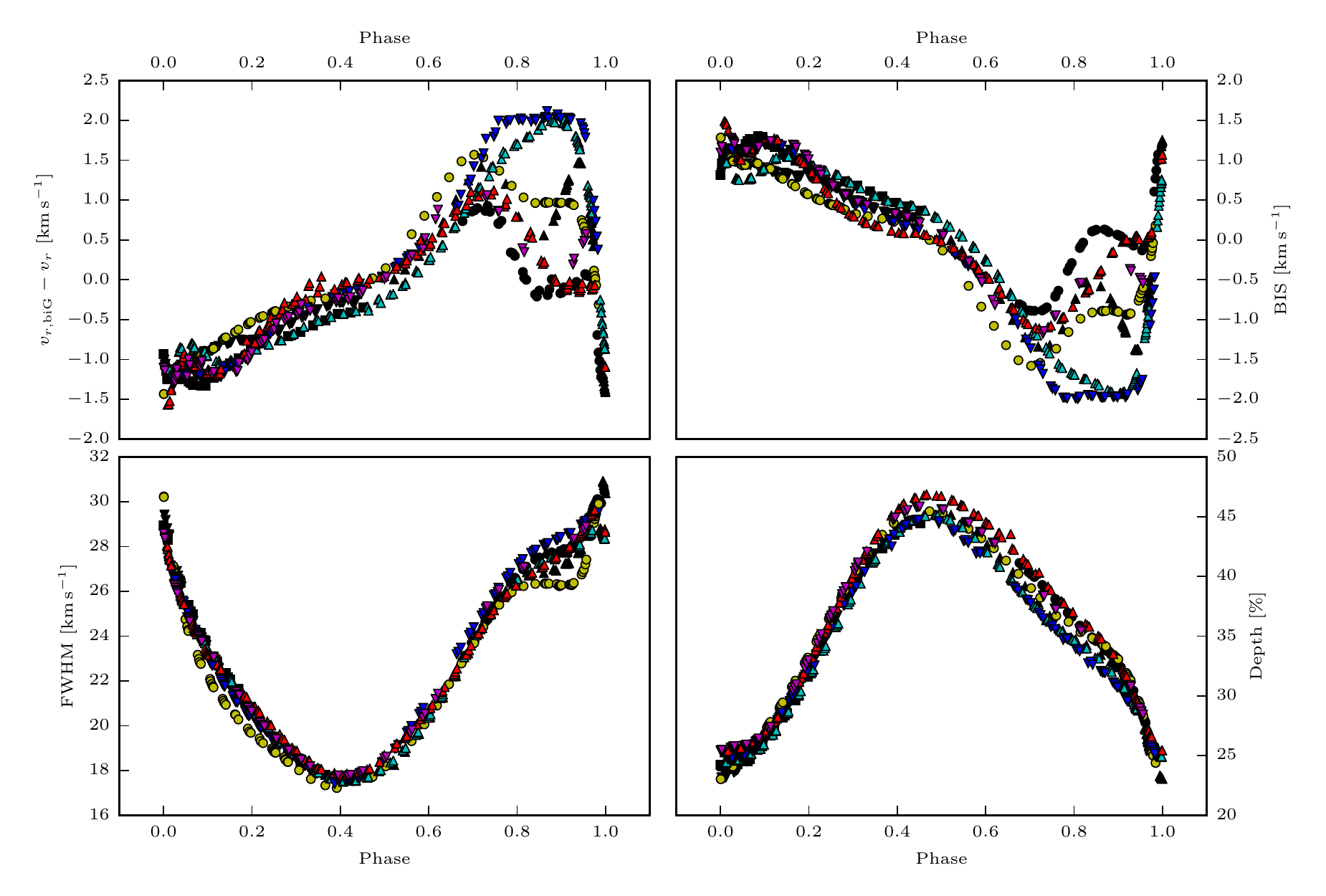}
\caption{Variation of CCF shapes with pulsation phase. {\it Top left:}
Difference of bi-Gaussian and Gaussian RV, {\it top right:} Bisector
inverse span (BIS), {\it bottom left:} FWHM of fitted Gaussian, {\it bottom right:} Depth in
percent of fitted Gaussian.}
\label{fig:LineShapePhase}
\end{figure*}

Four quantities are used to describe CCF variability:
1) the difference between bi-Gaussian and Gaussian RVs ($v_{r,\rm{biG}} - v_r$); 2) the bisector
inverse span (BIS), cf. Sec.\,\ref{sec:obs:CCFs+RVs}; 3) FWHM, the full width
at half-maximum of the fitted Gaussian; 4) normalized CCF depth, i.e., the normalized peak
height (here computed as a depth to resemble an absorption line) of the CCF. Figure\,\ref{fig:LineShapePhase} illustrates the variation of these parameters
as a function of phase and their  modulated (cycle-to-cycle and longer-term)
character.

\begin{figure}
\centering
\includegraphics{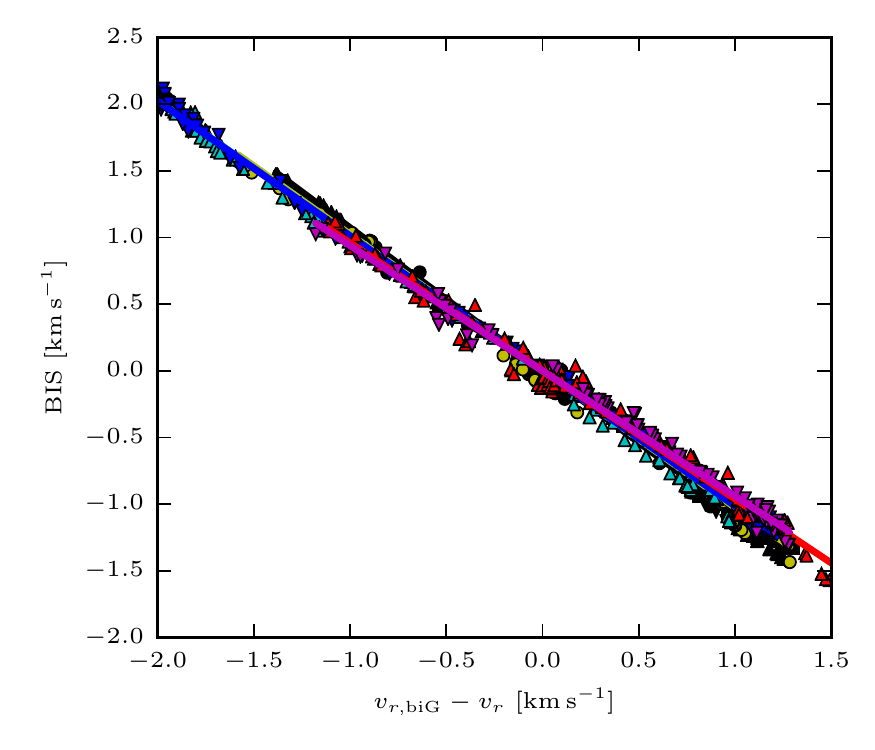}
\caption{Bi-Gaussian RV offset against bisector inverse span (BIS)
determined from the CCF. Each cycle is fitted separately, the average slope
over all cycles is $-1.009$.}
\label{fig:BISvsBiGauss}
\end{figure}

The difference between bi-Gaussian and Gaussian RVs based on a G2 mask varies
between approximately $-1.7$ and $2.1$\,\kms.
Conspicuously, this difference is opposite to the BIS variation and has nearly
identical amplitude. This correspondence is a consequence of the bi-Gaussian's
construction as an asymmetric line profile.
BIS is defined such that it is negative when the line core is red-shifted
compared to the higher sections of the CCFs, ergo the bi-Gaussian RV for a
negative BIS is larger than the Gaussian RV, and vice versa for positive BIS
(core more blue-shifted than upper CCF regions).
As Figure\,\ref{fig:BISvsBiGauss} shows, there is a near one-to-one
correspondence between $v_{r,\rm{biG}} - v_r$ and the BIS parameter for all
pulsation cycles. Linear regressions assuming a fixed intercept at (0,0) yield
slopes between -0.94 and -1.07 for the individual pulsation cycles, whereas a
regression to all data has slope $-1.009$.

\begin{figure*}
\centering
\includegraphics{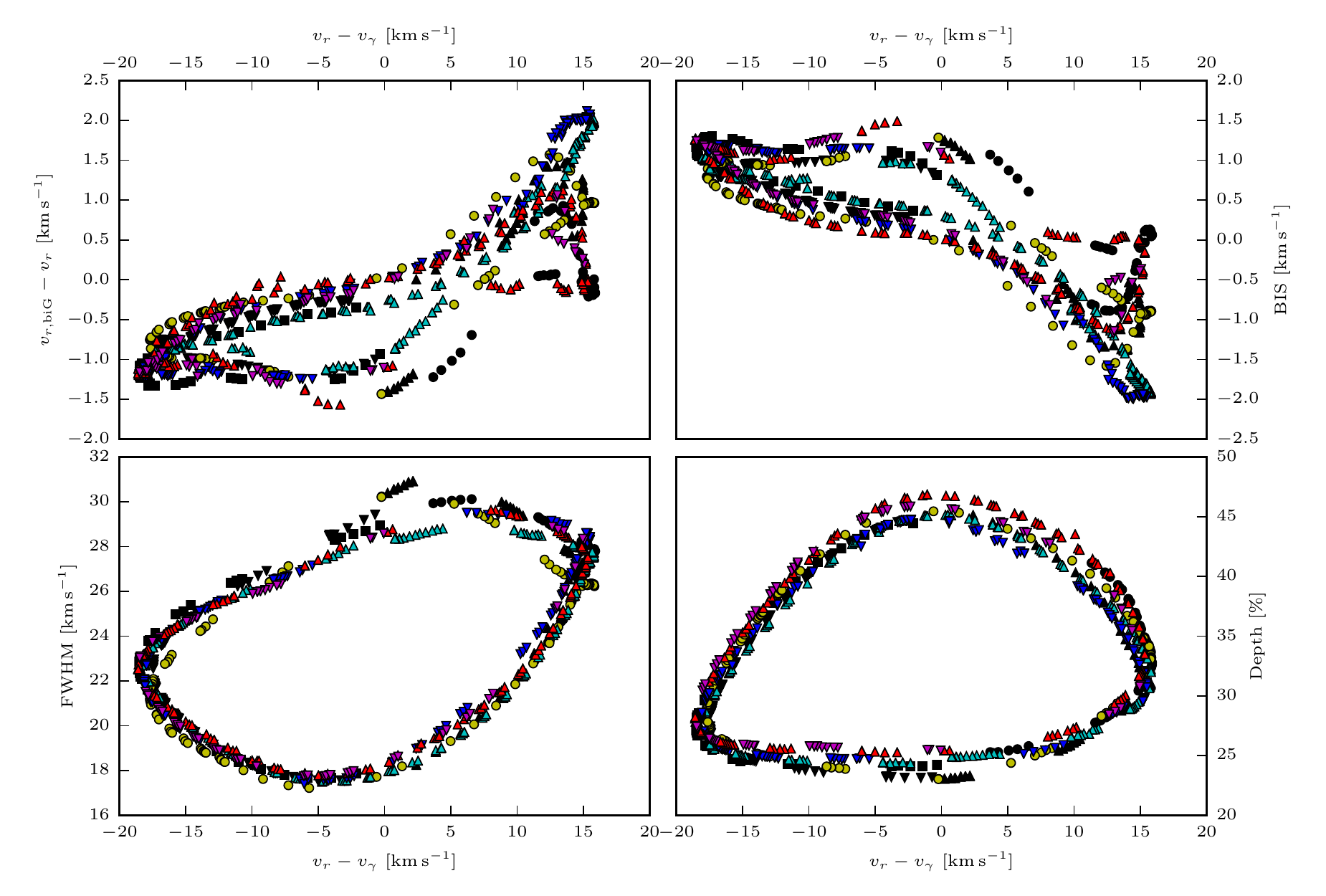}
\caption{Variation of CCF line shape parameters against RV, shown here centered
on the pulsation-averaged velocity $v_\gamma$.
{\it Top left:} bi-Gaussian minus Gaussian RV, {\it top right:} Bisector
inverse span (BIS), {\it bottom left:} FWHM of fitted Gaussian, {\it bottom right:} Depth in
percent of fitted Gaussian.}
\label{fig:LineShapeVrad}
\end{figure*}

The primary origin of  spectral line asymmetry in Cepheids are
rotation \citep{2007PASP..119..398G} and velocity fields 
\citep[e.g.][]{1975ApJ...201..641K,2006A&A...453..309N}. The rotational
contribution to line asymmetry originates in the convolution of the rotation profile and the 
pulsation velocity. Thus, the rotational contribution to BIS 
is strongest, when the pulsation velocity is extreme (minimal or maximal), 
and this overall pattern is clearly observed in the BIS parameter. 

The surface rotation velocity $v_{\rm{eq}}$ of a Cepheid is expected to vary by
up to $10\,\%$ over the course of an expansion-contraction half-cycle due to
conservation of angular momentum \citep{2007PASP..119..398G}. 
However, since the greatest difference in $v_{\rm{eq}}$ occurs 
at minimum and maximum radius where the pulsational velocity vanishes,
no significant contribution to BIS is expected due to this effect.

The contribution of pulsation-induced velocity fields to line asymmetry can be
understood as the added contributions of line forming regions moving at
different velocities. For instance, if higher layers are expanding more quickly
than lower ones, positive BIS is to be expected. Conversely, negative BIS is
expected when deeper layers are contracting more slowly than higher layers. The
steep sign reversal near minimal radius ($\phi \sim 1$) is thus a consequence of
the outward-directed shock wave reversing the velocity gradient. These
relationships are further investigated in \S\ref{sec:obs:RVgrad}.

In (non-variable) cool supergiants, velocity fields due to granulation  create
significant line asymmetry, with bisectors exhibiting a smooth variation as
function of spectral type. Full bisector velocity spans (top minus bottom) range
from $\sim 300 \hbox{--} 500$\,\ms\ near G4 (less at K2) up to $\sim -1$ to
$-2$\,\kms\ near F5\hbox{--}F8 \citep{1986PASP...98..499G}.
Intriguingly, the granulation-induced bisector asymmetry of non-variable
supergiants has opposite sign from the BIS variations shown in
Fig.\,\ref{fig:LineShapePhase}, even though $\ell$\,Car's variability spans
similar spectral types.
BIS is positive when $\ell$\,Car is hottest (maximal RV), and BIS is negative
when it is coolest (after maximal radius).
Thus, it appears that granulation is not the likely origin of $\ell$\,Car's
observed line asymmetry.

The variation of parameter BIS in Figure\,\ref{fig:LineShapePhase} exhibits
several noteworthy peculiarities. First, the steep rise immediately before phase
$1.0$ is the most consistent part among the different pulsation cycles. The
outward-directed pulsation wave initiated by the main pulsation mechanism (the
He\,II partial ionization zone) is undoubtedly responsible for this feature,
since it occurs very close to minimal radius. The occurrence of this significant
realignment further supports the choice of $v_\gamma$ as reference point for
timing the duration of individual pulsation cycles (\S\ref{sec:timings}).
Second, BIS reaches more extreme values during contraction ($\phi \sim 0.8
\hbox{--} 0.9$) than during expansion ($\phi \sim 0.1 \hbox{--} 0.2$) during
some pulsation cycles (c15a and b).
Third, while cycle-to-cycle differences in BIS are evident at all phases, they
are most strongly pronounced during contraction and exhibit a wave-like pattern
that differs from cycle to cycle, is not in phase with the pulsation, and
carries over into subsequent pulsation cycles. Thus, the atmospheric velocity
field during a given cycle retains a memory of the previous cycle.
Fourth, BIS modulation is more noticeable on longer timescales than among
subsequent cycles, possibly suggesting
(semi-)periodicity, given that the 2014 and 2016 cycles are more similar to each
other than to the 2015 cycles. A periodicity of this timescale ($\sim 2$ years)
would be consistent with the order of magnitude expected for the rotational
period of a $\sim 180\,R_\odot$, $9\,M_\odot$ Cepheid
\citep{2004ApJ...604L.113K,2014A&A...564A.100A,2016A&A...591A...8A}.

The modulated variability of the BIS
parameter provides a crucial insight into the origin of RV curve modulation. As explained in Sec.\,\ref{sec:obs}, RV
measurements obtained by fitting Gaussian profiles to CCFs are biased 
\citep{1982A&A...109..258B}.
However, if the shape of spectral lines at a fixed phase were to repeat
perfectly, then Gaussian RVs would be subject to the same bias in each pulsation
cycle and thus yield consistent results at fixed phase.
The modulated BIS variability here discovered demonstrates that line shapes are
not consistent between pulsation cycles, thus leading to RV curve modulation.
Bi-Gaussian RVs are even more strongly affected by these
cycle-to-cycle differences in line shape, as expected due to their asymmetric 
construction (cf. Figs.\,\ref{fig:RVmodulationZoomin} and
\ref{fig:BISvsBiGauss}). This link between line asymmetry,
velocity gradients, and RV curve modulation is explored in detail using a
Doppler tomographic method in \S\ref{sec:RVgrad}.

Figure\,\ref{fig:LineShapePhase} further reveals peculiar differences in the
FWHM and CCF depth parameters among pulsation cycles. As
was the case for BIS, the most conspicuous differences are seen near pulsation phase $0.9$.
Line width at these phases has previously been discussed in terms of 
shock-induced turbulence
\citep[e.g.][]{1973ApJ...180..895K,1996A&A...307..503F}.
The cycle-to-cycle variations of the FWHM parameter do not directly correlate with
those exhibited by BIS, cf. yellow circles and blue downward triangles, for
instance.

Figure\,\ref{fig:LineShapeVrad} complements Fig.\,\ref{fig:LineShapePhase} by
illustrating the variability of CCF shape parameters BIS, FWHM, and CCF depth as
a function of $v_r$, centered on $v_\gamma$. In particular, this reveals
butterfly-shaped diagrams for BIS (and thus also the $v_{r,\rm{biG}} - v_r$
difference). Figure\,\ref{fig:LineShapeVrad} features smooth variations that
differ noticeably among the pulsation cycles. It clearly shows that BIS tends to
be most extreme when RV is extreme, which illustrates the significant impact of
surface rotation on line asymmetry. An important exception to this general
behavior is the peak with positive BIS near the top center of the plot, during
expansion. This phase coincides with the piston phase of the pulsation, when
velocity gradients are expected to be strongest due to the outward-directed
pulsation wave passing through the atmosphere. The variation of FWHM is 
also most disturbed near minimum radius ($v_r \sim v_\gamma$ and maximal FWHM).

\begin{figure*}
\centering
\includegraphics{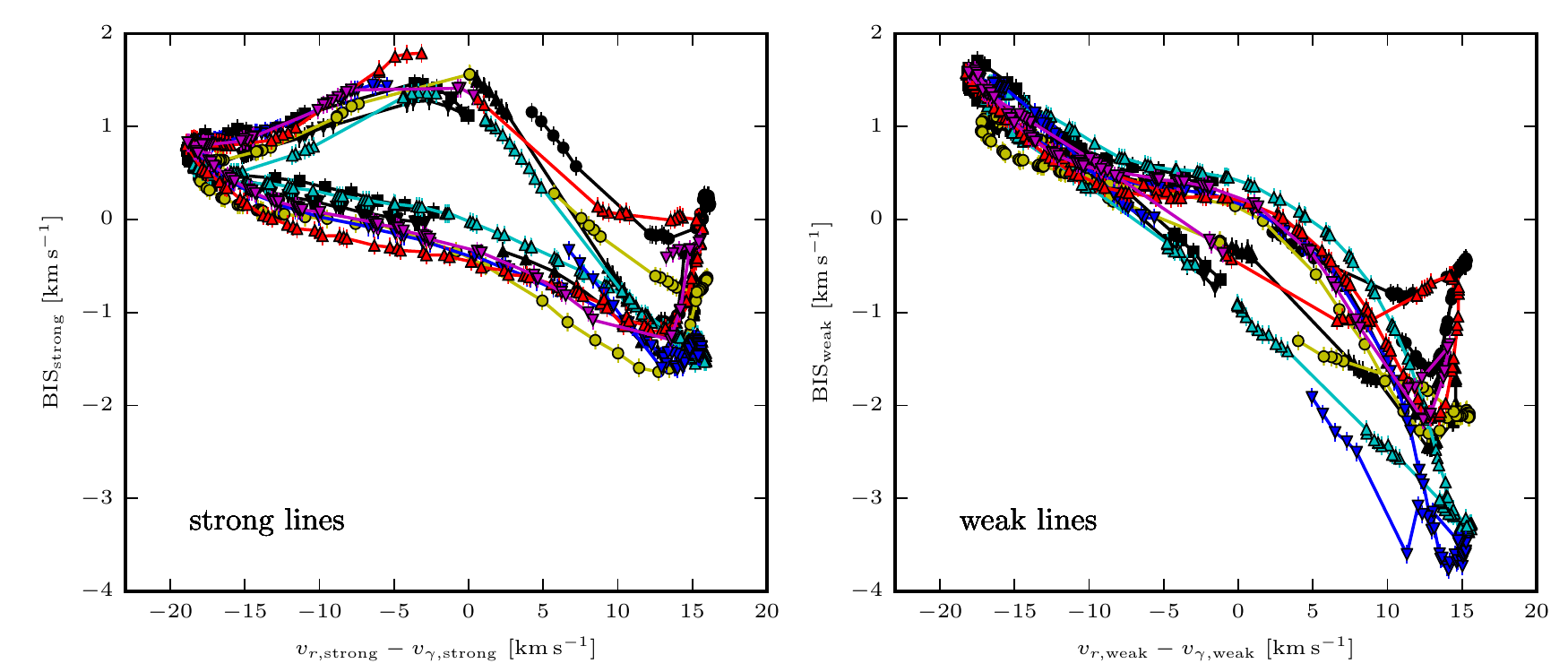}
\caption{BIS against RV for strong-line CCFs (left-hand panel) and weak-line
CCFs (right-hand panel). Weak lines exhibit a broader range of asymmetry than
strong lines and are subject to stronger modulations near maximal velocity
(fastest contraction).}
\label{fig:RVvsBIS_highmask+lowmask}
\end{figure*}

\begin{figure*}
\centering
\includegraphics[]{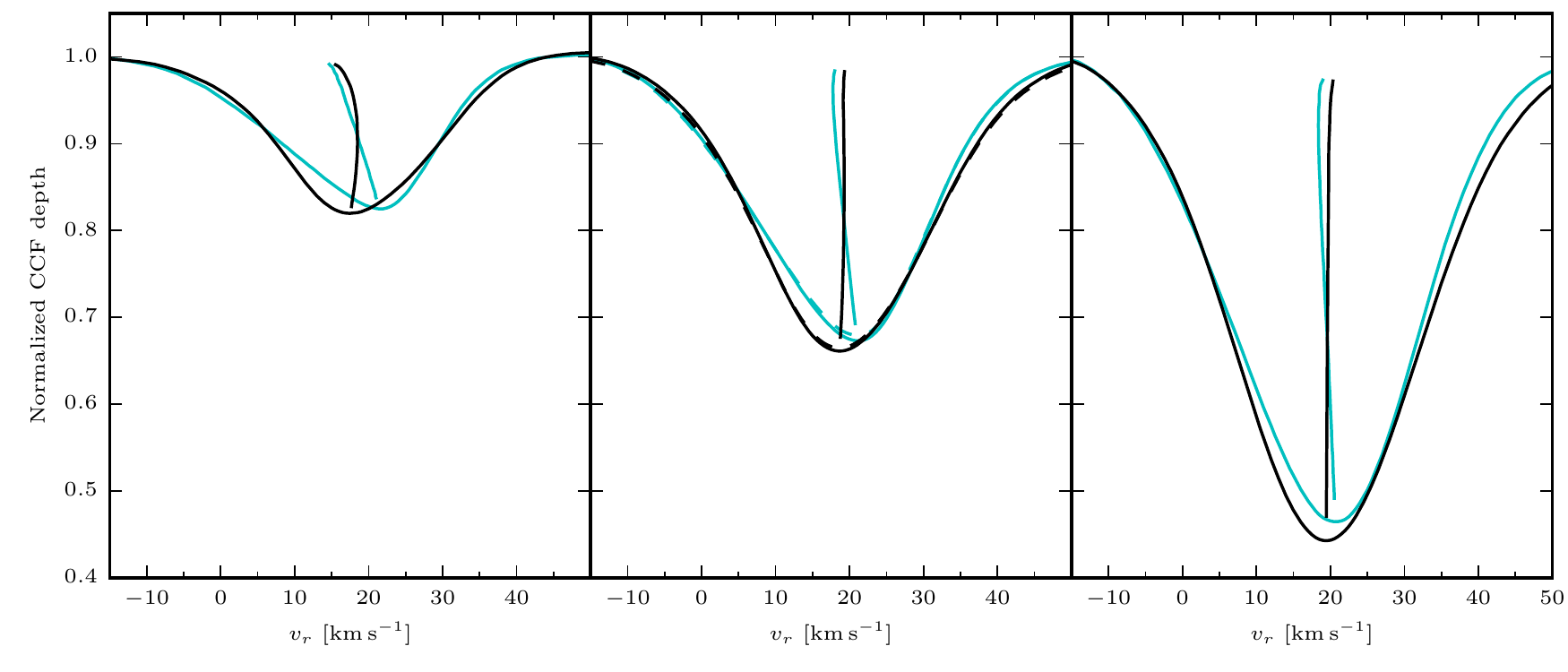}
\caption{Difference in CCF shape at nearly identical phase as observed in c15a
($\phi = 0.869$, cyan) and c16a ($\phi = 0.876$, black) where the difference in
BIS is maximal among pulsation cycles, cf. Fig.\,\ref{fig:LineShapePhase}.
Left-hand panel shows weak-line CCFs, right-hand panel strong-line CCFs.
Center panel shows CCFs based on G2 mask as solid lines as well as the summed
weak and strong-line CCFs as dashed line. Line bisectors are shown
for each mask computed.}
\label{fig:CCFdifferencesFixedphase}
\end{figure*}

Figure\,\ref{fig:RVvsBIS_highmask+lowmask} shows the computed BIS and RVs for
both the weak and strong-line CCFs, in analogy with the upper right panel of
Fig.\,\ref{fig:LineShapeVrad}. The resulting BIS loops are significantly
different for strong and weak line CCFs. Several interesting features appear in
these diagrams. Firstly, BIS$_{\rm{weak}}$ variations reach much more extreme
negative values at maximum velocity than BIS$_{\rm{strong}}$. This is further
illustrated by Fig.\,\ref{fig:CCFdifferencesFixedphase}.
Secondly, observed cycle-to-cycle differences in BIS are largest at maximum RV,
just as the observed differences in $v_r$.
Thirdly, BIS$_{\rm{weak}}$ exhibits more intense modulation than
BIS$_{\rm{strong}}$, which also matches the RV curve modulation seen in
Fig.\,\ref{fig:RVmodulationZoomin}.
For strong line CCFs, these loops differ more strongly at every point along the
duty cycle compared to weak line CCFs, although the differences at the extremes
is less pronounced.
For instance, c15a yields a much more compact loop than c16b. Strong line loops
(BIS$_{\rm{strong}}$ vs. $v_{r,\rm{strong}}$) are much more open during
expansion than weak line loops, for which the opposite appears to be the case.

The loops in Figure\,\ref{fig:RVvsBIS_highmask+lowmask}
exhibit the widest opening when $v_r \sim v_\gamma$, i.e., close to times of
extremal radius. The closer and steeper correlation  between BIS$_{\rm{weak}}$
and $v_{r,\rm{weak}}$ compared to the same quantities derived for strong line CCFs
is peculiar and may be indicative of depth-dependent differential
rotation, since line asymmetry is in large parts due to the convolution of the
rotation and pulsational velocities.

Besides rotation, velocity fields contribute significantly to line
asymmetry. The relationship between the metallic velocity gradient and BIS
parameter is therefore investigated in detail in the following
\S\ref{sec:RVgrad}.

\subsection{Velocity Gradients}\label{sec:RVgrad}

Cepheid atmospheres are highly dynamic and characterized by strong velocity
fields.
Previous work has shown in detail the different velocities exhibited at fixed
phase for spectral lines belonging to different elements, ionization potentials,
and line depths 
\citep[e.g.][]{1956ApJ...123..201S,1978ApJ...222..578K,1989ApJ...337L..29S,1990ApJ...362..333S,1992MNRAS.259..474W,2015PASP..127..503W,1993ApJ...415..323B,2006A&A...453..309N,2007A&A...471..661N,2009A&A...507..397H}

The metallic line velocity gradient, $\delta v_r (t)$, is computed here as the
difference between strong-line and weak-line RVs as defined in
Eq.\,\ref{eq:rvgradient} to investigate the relationship between the
BIS parameter and the atmospheric velocity field. The key benefit of the
presently used definition of $\delta v_r (t)$ is its ability to reveal 
even small differences (both at the same time and between different pulsation
cycles) thanks to the very high RV precision afforded by the cross-correlation
technique.

However, the present approach is also limited by several choices, including
blurring due to the inclusion of many spectral lines formed at different heights
for each mask, the approximate nature of using spectral lines of different
strength to trace different atmospheric layers, and the possible influence of
different portions of the stellar disk being probed by different line masks.
Nevertheless, this approach is useful and tests performed using different line
selections revealed similar trends, albeit with different amplitudes.

\begin{figure*}
\centering
\includegraphics{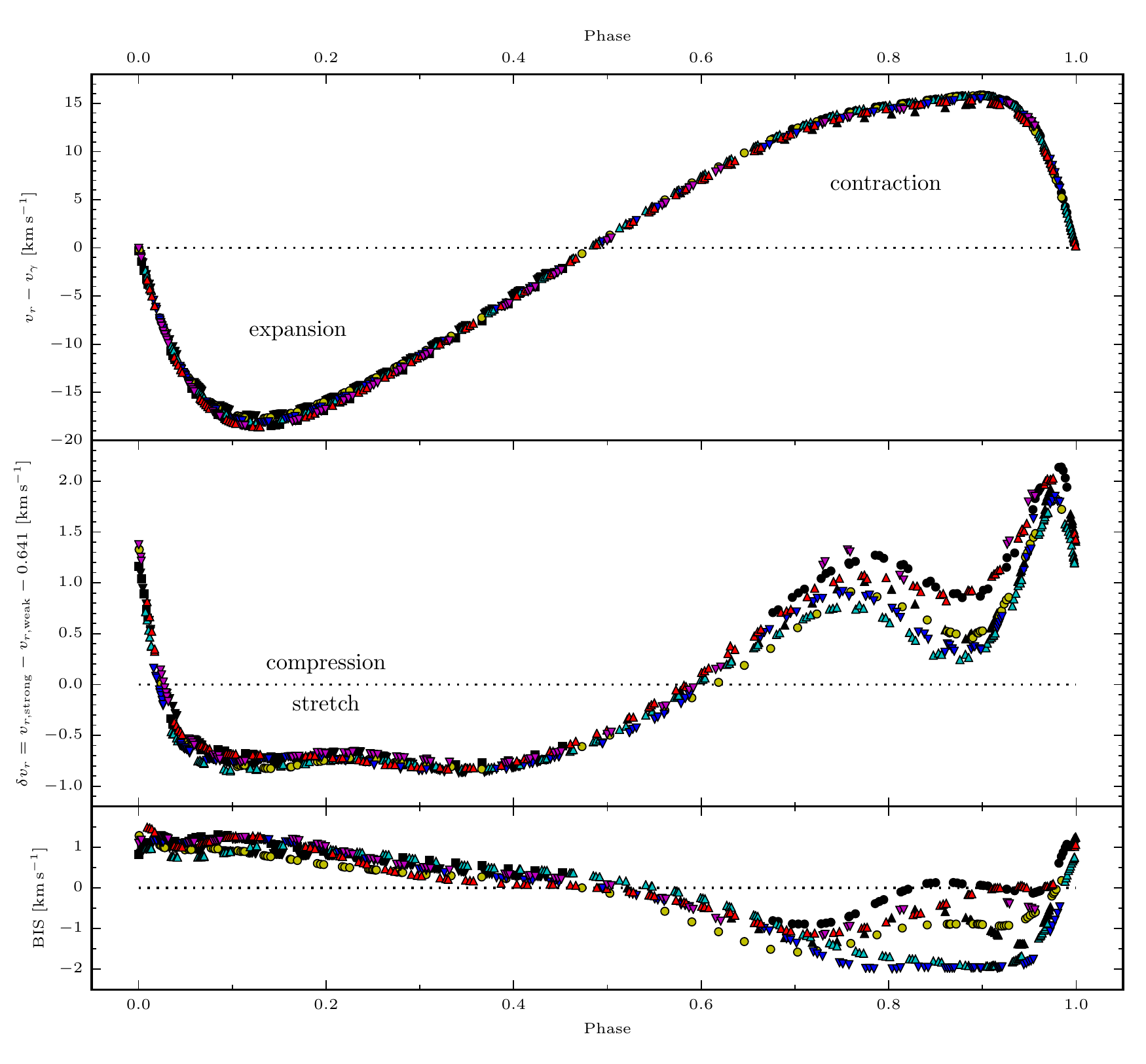}
\caption{Time-dependent velocity gradients $\delta v_r$ ({\it center}) and their
correspondence to the overall RV variability $v_r$ ({\it top}) and the bisector
inverse span (BIS, {\it bottom}). $\delta v_r > 0$ indicates compression, cf.
Tab.\,\ref{tab:DTexplained}. Velocity gradients and line asymmetry are clearly
correlated, both exhibit patterns that are more similar in 2014 and 2016 than in 2015.}
\label{fig:DTresidualBIS}
\end{figure*}

\begin{figure}
\centering
\includegraphics{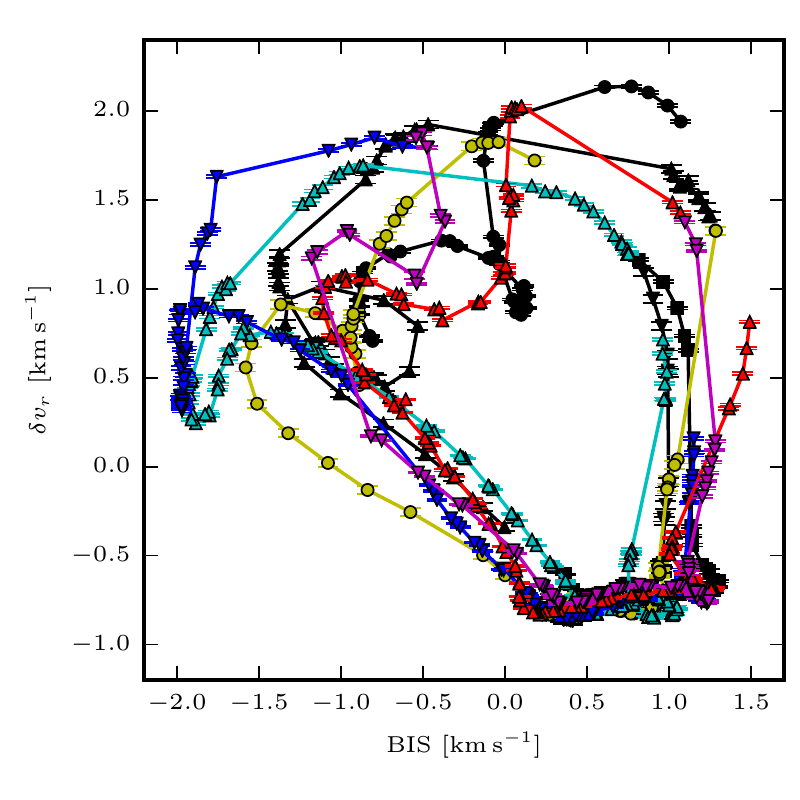}
\caption{Velocity gradient $\delta v_r$ against CCF asymmetry parameter BIS.
$\delta v_r$ corresponds more closely to BIS when $\delta v_r < 0$, i.e.,
when the atmosphere is being stretched.}
\label{fig:DTvsBIS}
\end{figure}

Figure\,\ref{fig:DTresidualBIS} illustrates $\delta v_r$ (center panel) against
pulsation phase together with the RV curves ($v_r$, top panel) and the BIS
variation (bottom panel). The dependence of $\delta v_r$ on BIS is further shown
in Fig.\,\ref{fig:DTvsBIS}.

Figure\,\ref{fig:DTresidualBIS} reveals a globally smooth variation of the
velocity gradient as a function of pulsation phase that is dominated by three
humps of increasing amplitude located at phases of approximately 0.25, 0.77, and 0.97. A first minimum before the first hump at phase 0.1 \hbox{--} 0.15 coincides with the
peak of the RV curve, i.e., where the shock wave emerges from below. The slight
hump indicating reduced stretch thus appears to be related to the trailing part
of this shock wave.

Shortly after maximum radius (at phase $\sim 0.6$), $\delta v_r$ turns over from
negative to positive values as the star is contracting.
This compression first proceeds in an accelerated fashion until a maximum is
reached near phase 0.77 and the compression is slowed temporarily before being
re-accelerated forcefully towards the highest peak near phase 0.97.
The variation of $\delta v_r$ between phase 0.75 and 0.95 (well before minimum
radius is reached) is indicative of a process beyond gravitational collapse that
first works against the acceleration of compression and subsequently contributes
to it. This is certainly a feature of the pulsation mechanism, given that it
is apparent during every pulsation cycle.

Finally, near phase 0.95-0.98, $\delta v_r$ is reminiscent of a
discontinuity and experiences a sharp turnaround, proceeding from maximal
compression to maximum stretch in less than $10\%$ of a pulsation period. This
feature coincides with the very fast decrease in RV, cf.
Fig.\,\ref{fig:RVmodulationZoomin}, which initiates the expansion.

Figure\,\ref{fig:DTresidualBIS} also reveals the existence of significant
cycle-to-cycle changes in $\delta v_r$. The most striking  differences among
pulsation cycles take place during contraction, start near $\phi=0.6$, and
proceed until expansion is initiated. This behavior is analogous to the greater
extent of RV curve modulation observed during contraction (near maximum RV).
Smaller cycle-to-cycle differences near minimum RV (phase 0.1 \hbox{--} 0.15)
further coincide with lower RV curve modulation at minimum $v_r$. $\delta v_r$
also exhibits more consistent behavior for intermediate phases during which
little to no cycle-to-cycle $v_r$ modulation is found.

The smooth variation of the each cycle's $\delta v_r$ curve demonstrate that the
present approach traces even small changes in velocity gradients with
confidence.
Similarly to the behavior of RV curve modulation, consecutive
pulsation cycles tend to reproduce similar $\delta v_r$ variability, whereas
longer timescales lead to larger differences.
Differently from the RV curve (Fig.\,\ref{fig:RVmodulationZoomin}), however, the
2014 and 2015 cycles are more similar to each other than to the cycles observed
in 2016. These different modulation patterns among the various spectral
indicators suggest a highly complex behavior of modulated line profile
variability across all spectral lines.

Comparing cycle-to-cycle changes in $\delta v_r$ with those observed for BIS in
the bottom panel of Fig.\,\ref{fig:DTresidualBIS} shows both indicators to
correlate very closely. For instance, close to phase 0.1 (near fastest $v_r$),
c15a exhibits both lowest BIS and lowest $\delta v_r$, whereas c16b exhibits
greatest BIS and $\delta v_r$. Close to phase $0.8$, c16a yields the highest
value for $\delta v_r$ and BIS, and similarly, c15b exhibits minimal BIS and
$\delta v_r$. However, the correspondence is not perfect and some smaller
differences remain. The need to adopt a common value for $v_\gamma$ for all
pulsation cycles for timing purposes (cf. \S\ref{sec:timings}) likely
dominates these small differences.
Further reasons for non-correspondence include the fact that BIS is measured on
CCFs computed using the G2 mask, which contains some spectral lines not included
in the strong and weak line masks (depth range 0.55 \hbox{--} 0.65). However,
this contribution is expected to be small, since the addition of weak and
strong-line CCFs closely resembles G2 mask-based CCFs (cf.
Fig.\,\ref{fig:CCFdifferencesFixedphase}).

Figure\,\ref{fig:DTvsBIS} illustrates another important feature, namely that 
variations in $\delta v_r$ and line asymmetry
(BIS) correspond more closely during expansion ($\delta v_r < 0$) than during
contraction ($\delta v_r > 0$). This points towards a 
resetting effect of the outward-directed pulsation wave. Other atmospheric effects such as
convection and turbulence appear to cause greater departures from this
correspondence once the shock wave has passed through the atmosphere.

\begin{figure*}
\centering
\includegraphics{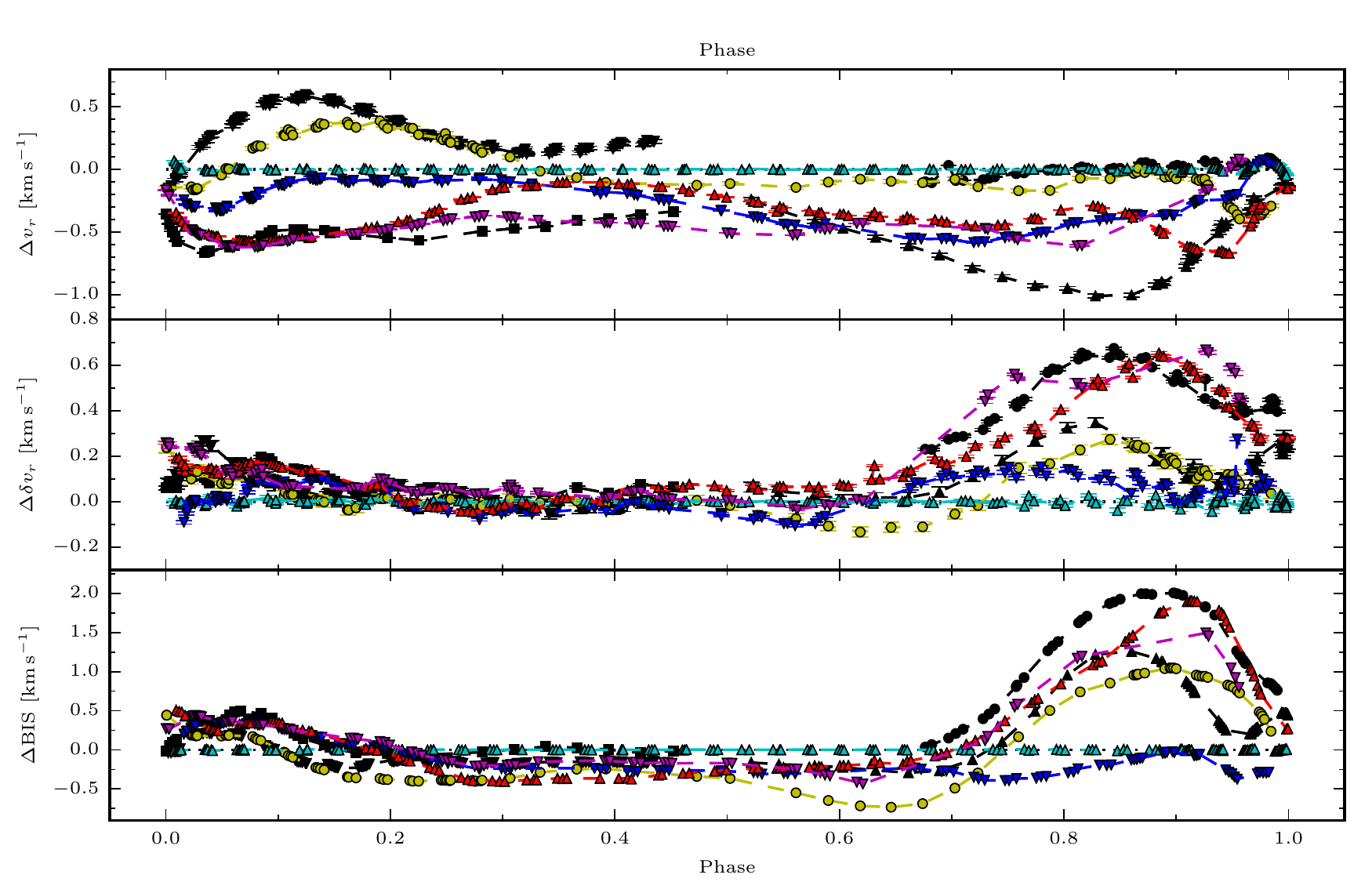}
\caption{Cycle-to-cycle and longer-term changes in RV, $\Delta v_r$ (top),
velocity gradient, $\Delta \delta v_r$ (center), and CCF asymmetry parameter
$\Delta$\,BIS (bottom) shown relative to c15a (cyan upward triangles).
Differences among cycles seen in $\delta v_r$ and BIS correlate closely,
indicating that BIS is a suitable proxy to trace velocity gradients.
Changes in $v_r$ do not mirror directly the changes in
velocity gradient or BIS.}
\label{fig:DeltaDTDeltaBISvsPhase}
\end{figure*}

\begin{figure}
\centering
\includegraphics{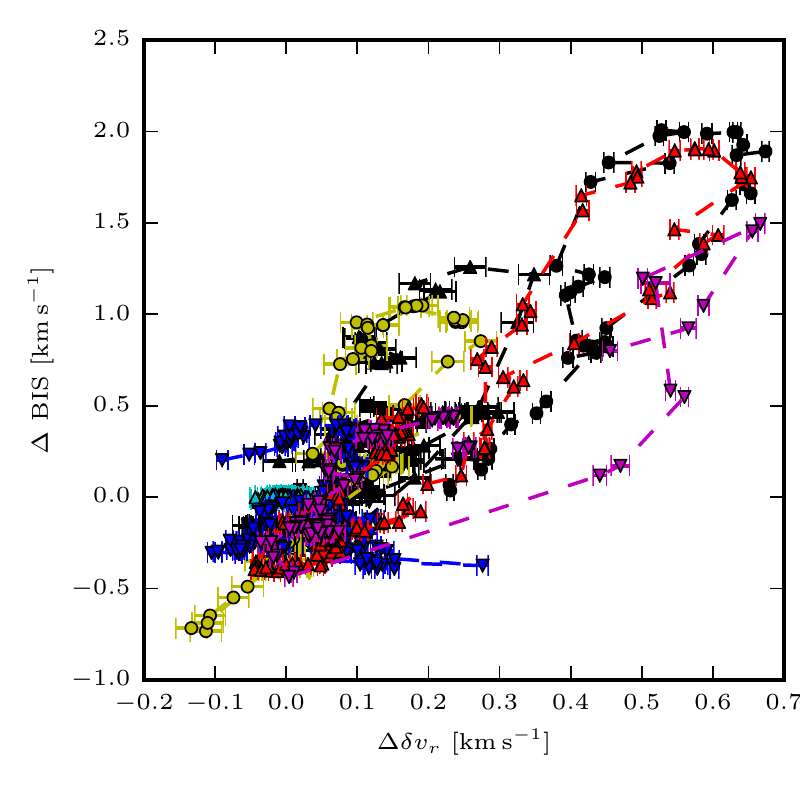}
\caption{Cycle-to-cycle changes in velocity gradient $\delta v_r$ against those
exhibited by BIS calculated relative to Fourier series fit to c15a data.}
\label{fig:DeltaDTvsDeltaBIS}
\end{figure}

Figures\,\ref{fig:DeltaDTDeltaBISvsPhase} and \ref{fig:DeltaDTvsDeltaBIS} serve
to further illustrate the correspondence between modulated BIS and $\delta v_r$.
Figure\,\ref{fig:DeltaDTDeltaBISvsPhase} shows this behavior as a function of
phase, whereas Fig.\,\ref{fig:DeltaDTvsDeltaBIS} plots $\Delta \delta v_r$
against $\Delta$BIS. It is worth recalling here the nomenclature adopted where
$\delta$ indicates the difference between measurements obtained at the same time
and $\Delta$ denotes differences between pulsation cycles.  Both figures use
pulsation cycle c15a as a reference to compute cycle-to-cycle differences.
The amplitude of $\Delta$BIS is approximately a factor of 3
larger than the amplitude of $\Delta \delta v_r$, most likely due to the
difference in spectral lines between the G2 mask used to compute BIS and the
fewer and different lines used to compute weak and strong line CCFs.  For
$\Delta \delta v_r$ the maximal difference among 2016
and 2015 pulsation cycles reaches  $\sim 700$\,\ms, which is on the order of
$50\%$ of the average $\delta v_r$ at the same phase ($\phi \sim 0.875$).
For BIS, cycle-to-cycle differences reach up to $\sim 2.1$\,\kms\ at this phase,
with BIS ranging from $+0.1$ to $-2.0$\,\kms.

The correspondence between the shape of the $\Delta \delta v_r$ and $\Delta$BIS
curves against phase (Fig.\,\ref{fig:DeltaDTDeltaBISvsPhase}) is striking and
demonstrates that cycle-to-cycle and longer-term changes in the velocity
gradient are primarily responsible for the observed modulated BIS variability.
This important result suggests that cycle-to-cycle RV curve modulation in
long-period Cepheids discovered by A14 is primarily due to cycle-to-cycle and
longer-term variations in velocity gradients that modify the spectral line
variability, acting primarily on line asymmetry.
Despite a remarkable correspondence between the parameters,
Fig.\,\ref{fig:DeltaDTDeltaBISvsPhase} also shows that $\Delta v_r$ does not
correlate immediately with changes in the velocity gradient ($\Delta \delta
v_r$).
This is likely due to the combined influence of line profile variations on FWHM,
depth, and BIS, of which only BIS was considered here.  
Further research is required to test whether such correlations could be used to
reduce the impact of modulated line profile variability on Cepheid RV
measurements.

Since cycle-to-cycle changes in velocity gradient correlate with those in BIS,
it follows that BIS can serve as a suitable proxy to detect changes in velocity
gradients. This is particularly useful to extend the investigation of
cycle-to-cycle differences to fainter stars, for which the here presented
Doppler tomographic technique cannot be applied due to insufficient spectral
SNR.
Specifically, investigation of BIS can help to distinguish between time
variations in the pulsation averaged velocity $v_\gamma$ due to
spectroscopic binarity (no cycle-to-cycle difference in BIS) and modulated line shape
variability. This is useful to detect companions with low mass ratios
\citep[such as $\delta$\,Cep's spectroscopic companion, see][]{2015ApJ...804..144A}.

\section{Discussion}\label{sec:disc}
The above results reveal a first insight into the highly complex modulated line
profile variability of Cepheids. 
This work has focused on quantities accessible via CCFs in order to start
exploring the origin of RV curve modulation discovered recently (A14).
Work in progress will expand this investigation to individual spectral lines. It
should be noted that RV curve modulation in short and long-period Cepheid occurs
on different timescales. Hence, the following discussion of the above results
should apply primarily to Cepheids with pulsation periods on the order
of $20 \hbox{--} 60$\,d. Further research is required to investigate how these
phenomena translate to other Cepheids and/or period ranges.

\subsection{On CCF asymmetry and velocity gradients}\label{disc:gradients}

The preceding sections have revealed a highly complex and cycle-dependent line
profile variability.
Although this paper is purely observational, it is useful to summarize the
key results capable of informing the  astrophysical interpretation of the
observed modulations. The most important clues found here include (in arbitrary
order):
\begin{enumerate}
  \item pulsation cycles retain memory of the preceding cycle;
  \item long-term modulation tends to dominate over short-term modulation;
  \item there is tentative evidence for a repetition in BIS modulation (2014
  compared to 2016 data);
  \item $\Delta$BIS correlates closely with $\Delta \delta v_r$;
  \item the strongest cycle-to-cycle differences in $\delta v_r$ occur during
  contraction, well before minimal radius;
  \item the asymmetry of weak lines more strongly correlates with pulsational
  velocity than for strong lines;
  \item weak lines show greater asymmetry and exhibit greater
  cycle-to-cycle modulation than strong lines.
\end{enumerate}

Changes in the velocity field could be explained by several mechanisms,
including convection, additional (e.g., non-radial or strange) pulsation modes,
surface inhomogeneities (spots), or inelastic shock. Granulation has previously
been mentioned as a possible explanation for so-called ``period-jitter'' in
Cepheids \citep{2014A&A...563L...4N}. As mentioned in \S\ref{sec:obs:RVgrad}
however, granulation-induced line asymmetry in non-variable supergiants has
opposite sign of $\ell$\,Car's line asymmetry.
Hence, other asymmetry-inducing effects such pulsation-induced velocity
gradients and rotation complicate the assessment of granulation based on line
asymmetries. 
Multi-dimensional models of pulsation-convection
interactions  \citep{2013MNRAS.435.3191M,2015MNRAS.449.2539M} should provide
interesting insights into this possible explanation. Convective perturbations of
the velocity field are likely to occur, since convective cells in cool
supergiants are large and $\ell$\,Car is a particularly cool Cepheid.
$\Delta$BIS is particularly strong when $\ell$\,Car is coldest, further
corroborating a link with convection.

Rotation and associated magnetic phenomena are an interesting possibility
primarily because of the tentative evidence of a pattern repeating over a 2-year
timescale, which is broadly consistent with the expected rotation period of
$\ell$\,Car. In addition, the dependence of line asymmetry on formation height
indicates possible differential (depth-dependent) rotation. While a magnetic
origin of amplitude modulations has been suggested \citep{2009ApJ...696L..37S},
little is known observationally about the magnetic fields of Cepheids \citep[see
e.g. $\eta$\,Aql in][]{2010MNRAS.408.2290G}. Inhomogeneities in the velocity
field of $\ell$\,Car's surface (e.g. due to spots) could lead to line asymmetries similar
to those observed in other rotating stars
\citep{1958IAUS....6..209D,1977SvAL....3..147G,1988ApJ...334.1008T}.

Shock associated with the pulsation has received much attention in the
literature
\citep[e.g.][]{1975ApJ...201..641K,1984ApJ...279..202S,1994ApJ...420..401B,2006A&A...457..575M,2014ApJ...794...80E,2016ApJ...824....1N}.
Unresolved, time-dependent line-splitting via the
\cite{1952SchwarzschildMechanism} mechanism could introduce line asymmetry,
which, if the shock were inelastic, could change from one cycle to the next and
affect the velocity field. Line splitting and emission of the shock-sensitive
Ca\,II K line in $\ell$\,Car was previously reported \citet{1969MNRAS.145..377D}
and is also seen at certain phases in the \Coralie\ spectra, together with
emission in H\,$\alpha$. Furthermore, UV emission observed at certain phases has
been linked to shock for $\ell$\,Car
\citep{1984ApJ...279..202S,1994ApJ...420..401B,2016ApJ...824....1N}. The strong
BIS modulation occurring during contraction could be related to such
inelasticity, or may be indicative of additional (e.g. higher-order
or non-radial) pulsation modes.

At present, these considerations remain of course speculative. Future work involving
additional stars, individual line profiles, and broader coverage of the
electromagnetic spectrum will allow additional insights into the complexity of
Cepheid pulsations.

\subsection{Implications for Cepheid RV measurements}\label{disc:RVmeas}

As shown in this paper, $\ell$\,Car's line profile variability (LPV) is subject
to modulation between consecutive pulsation cycles as well as on longer timescales.
Since RVs are measured via the Doppler shift of spectral lines, it is  
clear that RV curve modulation is a symptom of cycle-to-cycle changes in LPV.

In general, the concept of a single radial velocity is ill-defined for a
Cepheid due to well-known velocity gradients
\citep[e.g.][]{1956ApJ...123..201S,1969MNRAS.145..377D,1992MNRAS.259..474W,1993ApJ...415..323B,2007A&A...471..661N}.
While velocity gradients may bias individual RV measurements, they do not
preclude the recovery of the true pulsational variability if such bias
can be accounted for, e.g. via (phase-dependent) projection factors, see also
\S\ref{disc:pfactors}.
However, this work shows that the phase variability of velocity
gradients does not repeat perfectly between pulsation cycles, leading to a
complex time-dependence of RV variability. This problem is analogous to the difficulties encountered by
RV-based planet searches, where stellar signals due to activity or granulation
negatively impact the detectability of low-mass planets. Thankfully, 
the variability of the BIS parameter is a useful indicator for modulated LPV and
can help to distinguish between RV signals due to low-mass companions and
pulsation-related ``noise''.
At present it is however unclear how to mitigate the impact of modulated LPV on
RV measurements. 

The method for measuring RV (weak lines, strong lines, Gaussian,
bi-Gaussian) also impacts the resulting pulsation-averaged velocity $v_\gamma$,
cf. Tab.\,\ref{tab:DeltaRp}, limiting the ability to
search for low-mass companions using inhomogeneous data sets (cf. R.~I.~Anderson
et al submitted).
The dependence on $v_\gamma$ on line strength exceeds the dependence on the profile fitted to the computed CCF:
$v_\gamma$ differs by $\sim 600$\,\ms\ between weak- and strong-line RVs, 
whereas Gaussian and bi-Gaussian RV based on the G2 mask are consistent to
within a few tens of \ms. 
This behavior is certainly due to the stronger asymmetry of weak line CCFs
(\S\ref{sec:CCFs}), which is more pronounced near maximum than near minimum RV,
thus biasing $v_\gamma$ \citep[cf.][]{2008A&A...489.1255N}.
Studies aiming to investigate binarity or Galactic rotation curves
\citep[e.g.][]{1997A&A...318..416P} may thus benefit from employing strong-line
RVs, since these exhibit weaker asymmetry and are thus less biased. On the other
hand, studies interested in revealing modulated LPV in Cepheids may benefit from using
weak-line RVs as a first indicator.

In summary, different use cases may benefit from using differently defined RVs.
However, the definition of the RV measurement can lead to phase-dependent 
differences among RV measured using different instruments, or even by different
authors. Employing consistently defined RVs is thus crucial for high-precision
RV analyses, e.g. when investigating Cepheid binarity. For the time being, it is
unclear whether RV curve modulation can be avoided by defining the measurement
adequately. However, averaging over long temporal baselines may cancel out these
effects, cf. \S\ref{disc:pfactors} below.

\subsection{Implications for Baade-Wesselink Distance and p-factors}
\label{disc:pfactors}

RV curve modulation represents a difficulty for Baade-Wesselink type analyses
that exploit Cepheid pulsations to measure quasi-geometric distances.
Specifically, distance
\begin{equation}
d \propto \Delta R / \Delta \Theta = p / \Delta \Theta\
\int{v_r\, \rm{d}\phi}\ ,
\label{eq:BWdistance}
\end{equation}
where $p$ is the projection factor required to translate the
observed, disk-integrated line-of-sight velocity into the pulsational velocity,
$\Delta \Theta$ is the full-amplitude
angular diameter variation, and the RV integral is computed over the same phase
range. $\Delta \Theta$ can be averaged
over many cycles \citep[e.g.][]{2016A&A...587A.117B}, or determined for
individual half-cycles of expansion or contraction (A16). As argued in A14, cycle-to-cycle
and longer-term changes in RV amplitude and shape result in systematic
changes of the RV integral. This introduces a systematic distance uncertainty
if the measured $\Delta R$ and $\Delta \Theta$ are not equivalent, e.g. by not
being measured contemporaneously or by other systematics intervening even if measured
contemporaneously.

$p-$factors have been decomposed as follows \citep{2007A&A...471..661N}:
\begin{equation}
p = p_0 \cdot f_{\rm{grad}} \cdot f_{\rm{o-g}}\ ,
\label{eq:p-factor}
\end{equation}
where $f_{\rm{grad}}$ is a factor representing the impact of
velocity gradients, $f_{\rm{o-g}}$ represents the difference between the
motion of optical and  gas layers, and $p_0$ represents all other effects such
as geometry and limb darkening. 

A16 investigated whether angular diameter variations repeated perfectly, or
whether they, too, exhibit modulated variability. They further investigated
whether any modulation pattern would reproduce the trends exhibited by RV data.
The high-quality interferometric dataset obtained for $\ell$\,Car showed
tentative signs of modulated angular variability, although contributions from
instrumental effects could not be fully excluded.
Interestingly, however, RV and angular diameters exhibited very different
modulation behavior, which was interpreted as being the result of the different
motions of the optical continuum (measured by interferometry) and the gas
 (measured via spectral lines). This can be expressed as a complex time
 dependence\hbox{---}possibly changing from cycle to cycle\hbox{---}not
 previously considered for factor $f_{\rm{o-g}}$ in Eq.\,\ref{eq:p-factor}.

Section\,\ref{sec:RVgrad} demonstrates that $\ell$\,Car's velocity gradients
also exhibit a complex time-dependence, which is furthermore not in phase with
the pulsations and enters the definition of $p$ via a previously unknown
time-dependence of factor $f_{\rm{grad}}$. Having measured RV using different
line masks and profiles fitted to CCFs, let us now consider the impact of how RV
is defined on $p$.

Table\,\ref{tab:DeltaRp} lists values of RV integrals, $\int{v_r \rm{d}\phi} =
\Delta R/p$, computed for RVs based on three different correlation masks, as
well as bi-Gaussian RVs based on the G2 mask, measured for the individual
half-cycles accessible.
The duration of each half-cycle is determined as described in \S\ref{sec:obs}
using pulsation-averaged $v_\gamma$  (cf. \S\ref{disc:RVmeas}).
$\Delta R/p$ is then computed as done previously (A14,A16) using cubic splines
and Monte Carlo simulations (10\,000 draws). Note that durations and $\Delta
R/p$ can differ here from the values presented in A14 and A16 due to different
definitions of $v_\gamma$.

The row labeled $\langle \vert \Delta R/p \vert \rangle$ in
Table\,\ref{tab:DeltaRp} lists average absolute values of $\Delta R/p$ depending
on the definition of the RV measurement.
$\langle \vert \Delta R/p \vert \rangle$ depends significantly on the method
employed to measure RV (e.g. Gaussian vs. bi-Gaussian fits to CCFs), as expected
from the different RV amplitudes, cf.
Fig.\,\ref{fig:RVmodulationZoomin}. The row labeled $\sigma$ lists the standard
deviation of all $\vert \Delta R/p \vert$ and is followed by the fractional
standard deviation, $\sigma / \langle \vert \Delta R/p \vert \rangle$, which
shows that Gaussian RVs based on strong spectral lines yield the most consistent
result for $\Delta R/p$, exhibiting a scatter of $2.3\%$ compared to a scatter
of $3.0\%$ for bi-Gaussian RVs. Gaussian RVs based on the G2 mask yield the
second most consistent results among pulsation cycles, with a scatter of
$2.6\%$, followed by weak-line RVs. This behavior is directly related to the
cycle-to-cycle changes in the BIS parameter, which directly affects bi-Gaussian
RVs, and is expressed more strongly for weak-line CCFs, cf.
\S\ref{sec:CCFs}. 

Importantly, the definition of the RV measurement employed directly affects the
value of the projection factor obtained in empirical $p-$factor calibrations.
To illustrate this point, let us adopt $\ell$\,Car's distance of $497.5$\,pc
\citep{2007AJ....133.1810B}, the average $\langle \Delta \Theta \rangle =
0.569$ of two consecutive half-cycles (A16), and the average $\langle
\vert \Delta R/p \vert \rangle$ to determine $p = d \cdot \Delta \Theta / (
9.3095 \cdot \langle \vert \Delta R/p \vert \rangle) $, assuming $R_\odot =
696\,342$\,km \citep{2012ApJ...750..135E}. Statistical uncertainties are not included for this comparison whose aim is to illustrate the dependence of $p$ on
the RV measurement technique. $p$ is thus found to range from $1.207$ for
bi-Gaussian RVs to $1.330$ for weak-line RVs, even when averaging over many
pulsation cycles.
$p$ thus implicitly depends on the definition of the RV measurement by up to
$10\%$. This compares to the $\sim 10\%$ uncertainty on empirical $p-$factor
calibrations imposed by the accuracy of current parallax measurements
\citep{2016A&A...587A.117B}. 
As this comparison shows, employing a consistent definition of RV measurements
is crucial for determining $p-$factors and in particular
for calibrating a $p-$\Ppuls-relation
\citep{2007A&A...471..661N,2016A&A...587A.117B}. 

Summing over \drp\ of consecutive
half-cycles listed in Tab.\,\ref{tab:DeltaRp} reveals no significant net
growth or shrinkage in linear radius (mean growth is  $-0.04\,R_\odot/p$ with $\sigma=0.96\,R_\odot/p$ for Gaussian
RVs computed with G2 mask), i.e., short-term differences in $\Delta R/p$ cancel
out over longer timescales. This suggests that a consistent value of $\Delta
R/p$ can be determined if sufficiently many pulsation cycles are averaged.

\begin{table*}
\centering
\begin{tabular}{lrrrrrrrrr}
\hline
\hline
 & & \multicolumn{2}{r}{weak line, Gaussian} & \multicolumn{2}{r}{G2 mask,
 Gaussian} & \multicolumn{2}{r}{G2 mask, bi-Gaussian} &
 \multicolumn{2}{r}{strong lines, Gaussian} \\
 &  & \multicolumn{2}{r}{$v_\gamma = 2.930$\,\kms} &
 \multicolumn{2}{r}{$v_\gamma = 3.419$\,\kms} & 
 \multicolumn{2}{r}{$v_\gamma = 3.441$\,\kms} & 
\multicolumn{2}{r}{$v_\gamma = 3.571$\,\kms} \\
 Cycle & $N_{\rm{RV}}$
& duration & $\Delta R/p$ & duration & $\Delta R/p$ & duration & $\Delta R/p$ & duration & $\Delta R/p$ \\
 & & [d] & [$R_\odot$] & [d] & [$R_\odot$] & [d] & [$R_\odot$] & [d] &
 [$R_\odot$]
 \\
\hline
c14a & 55 & 18.467(7) & 22.135(8) & 18.379(7) & 22.888(7) & 18.13(2) & 24.47(2)
& 18.318(6) & 23.395(8) \\
c14b\_1 & 50 & 17.074(8) & -22.436(7) & 17.174(6) & -23.185(7) & 17.32(2) &
-24.35(2) & 17.250(6) & -23.830(7) \\
c14b\_2 & 37 & 18.555(8) & 23.489(8) & 18.442(6) & 24.102(8) & 18.37(2) &
26.10(2) & 18.363(6) & 24.556(8) \\
c14c$^\dagger$ & 115 & 16.84(6) & -22.02(1) & 16.96(6) & -22.74(1) & 17.3(2) &
-24.16(5) & 17.10(7) & -23.38(1) \\
c15a\_1 & 66 & 16.917(1) & -22.554(2) & 17.046(1) & -23.355(1) & 17.269(4) &
-24.950(3) & 17.133(1) & -24.006(1) \\ 
c15a\_2 & 81 & 18.624(1) & 23.848(2) & 18.500(1) & 24.374(1) & 18.256(4) &
26.509(3) & 18.417(1) & 24.802(1) \\ 
c15b\_1 & 86 & 17.087(2) & -22.873(2) & 17.223(1) & -23.686(1) & 17.328(4) &
-25.182(3) & 17.316(2) & -24.334(1) \\ 
c15b\_2 & 58 & 18.414(2) & 22.798(2) & 18.280(1) & 23.399(1) & 18.189(4) &
26.142(4) & 18.188(2) & 23.866(2) \\ 
c15c$^\dagger$ & 32 & 17.155(2) & -23.605(2) & 17.272(2) & -24.335(1) &
17.517(6) & -26.135(4) & 17.347(2) & -24.938(2) \\ 
c16b\_1 & 53 & 17.071(1) & -23.256(1) & 17.154(1) & -23.978(1) & 17.284(3) &
-25.274(3) & 17.214(1) & -24.580(1) \\ 
c16b\_2 & 57 & 18.420(1) & 22.551(2) & 18.340(1) & 23.430(1) & 18.237(4) &
24.338(3) & 18.283(1) & 24.151(1)  \\
c16c\_1 & 54 & 17.280(2) & -23.625(2) &
17.369(1) & -24.337(1) & 17.500(4) & -25.881(3) & 17.436(1) & -24.931(1)  \\
c16c\_2$^\ddagger$ & 19 & 18.277(2) & 21.986(4) & 18.183(1) & 22.962(2) &
18.043(4) & 23.967(7) & 18.121(2) & 23.724(3) \\
\hline
$\langle \vert \Delta R/p \vert \rangle$ & & & 22.860 & & 23.598 & & 25.189 & &
24.192
\\
$\sigma$ & &  & 0.648 & & 0.580 & & 0.885 & & 0.545 \\
$\sigma/\langle \vert \Delta R/p \vert \rangle$ & & & 0.028 & & 0.026 & &
0.030 & & 0.023 \\
\hline
\multicolumn{3}{l}{$p\ [ d=497.5\,\rm{pc},\ \Delta\Theta=0.56895\
\rm{mas}$]} &  1.330 & & 1.288 & & 1.207 & & 1.257
\\
\hline
\end{tabular}
\caption{Dependence of $v_\gamma$, half-cycle duration, integral of RV
curve\hbox{---}here denoted by $\Delta R/p$\hbox{---}, and projection factors
$p$ required for Baade-Wesselink distance measurements per pulsation half-cycle
(contraction/expansion) on measurement technique and lines used to compute RV.
Cycles are labeled as in Fig.\,\ref{fig:RVcurve}, with $\_1$ and $\_2$ denoting
first and second half of cycle. Statistical uncertainties are listed for
duration and $\Delta R/p$ using the notation $18.467(7) = 18.467 \pm
0.007$ and are based on 10\,000 Monte Carlo repetitions. 
$^\dagger$ marks cycles determined by extrapolation to nearest $v_\gamma$,
$^\ddagger$ marks the cycle with the fewest observations, for which the spline
fit is not as well constrained due to larger gaps in phase coverage.
Fluctuations of the average $\Delta R/p$ per method are $2\hbox{--}3$
percent.
$p$-factors are computed for each method assuming distance
\citep{2007AJ....133.1810B} and angular diameter variation \citep[using the
average of both measurements]{2016MNRAS.455.4231A} as stated to illustrate 
\emph{systematic} differences in $p$.}
\label{tab:DeltaRp}
\end{table*}

\section{Conclusions}\label{sec:summary}

This paper investigates the origin of cycle-to-cycle and longer-term modulations
of long-period Cepheid RV curves as discovered recently (A14) using
$\ell$\,Carinae as an example.
CCFs were computed based on $925$ high-SNR high-resolution
optical spectra observed during three campaigns (2014, 2015, and 2016), each of
which cover at least two complete consecutive pulsation cycles. 
Cycle-to-cycle differences in the spectral line profile
variability pattern are investigated and found to be significant, 
even among consecutive cycles, becoming 
more noticeable over longer timescales. The asymmetry parameter BIS exhibits the
most peculiar cycle-to-cycle variability and is considered in detail.

The dependence of the inferred RV variability on the measurement
technique is investigated by computing CCFs for three different line masks (G2,
weak lines, strong lines) and measuring RV by fitting either Gaussian or bi-Gaussian
profiles. Bi-Gaussian RVs exhibit stronger RV curve modulation than Gaussian
RVs, since the primary effect of modulated line profile variability concerns
asymmetry and since bi-Gaussians are by construction sensitive to such
asymmetry.
Weak-line CCFs generally exhibit stronger asymmetry than strong-line CCFs and
are more strongly affected by cycle-to-cycle variations.

Modulated BIS variability primarily originates in
long-term (cycle-to-cycle and longer) variations of atmospheric velocity
gradients and can therefore serve to identify this effect.
This is important, since BIS provides a straightforward means to distinguish
temporal variations in the pulsation-averaged velocity $v_\gamma$ due to modulated line profile
variability (modulated BIS variability) from ones caused by orbital motion (no
modulated BIS variability).

Visualization of cycle-to-cycle changes in velocity gradient indicates the
atmosphere to retain memory of the preceding cycle. The modulation
pattern seen for BIS suggests a possible repetition of this parameter's variability
over a timescale of $\sim 2$\,yr, which is close to the expected rotation
period for a star such as $\ell$\,Car.
Establishing a recurrence or even periodicity of this kind would be invaluable
for better understanding the origin of cycle-dependent velocity gradient
variations.

Possible origins of modulated line profile variations as well as their relevance
for BW distances are discussed, underlining the importance of
consistently defined RV measurements for BW distances. Since RV curve modulation
tends to average out over long timescales, it is advantageous for BW
analyses to reduce exposure to individual pulsation cycles. 
This will be of particular importance when {\it Gaia} will soon enable the
empirical calibration of projection factors for hundreds of Galactic Cepheids.

More generally, this work exposes previously unknown complexity in the
pulsation of Cepheids and opens a new window to further the understanding of
stellar pulsations.

\section*{Acknowledgments}
  Useful discussions with Xavier Dumusque and the assistance of many
  observers are acknowledged. I am grateful to the entire Euler team, the Geneva
  stellar variability group, and the Geneva
  exoplanet group for their assistance and support. The friendly
  and competent assistance by all ESO and non-ESO staff at La Silla Observatory was greatly appreciated. The anonymous referee's timely and positive response is acknowledged.
  
  The Swiss Euler telescope is funded by the Swiss National Science Foundation.
  The ability to operate such long-term campaigns on a small telescope with
  high-quality instrumentation was crucial to this work's ability to
  illustrate the complex behavior of Cepheid pulsations in this amount of
  detail.
  
  This research was funded by the Swiss National Science Foundation via an Early
  Postdoc.Mobility fellowship and has made use of NASA's ADS Bibliographic
  Services.


\bibliographystyle{mnras}
\bibliography{Bib_mine,Bib_Cepheids,Bib_modulation,Bib_DistanceScale,Bib_StellarRotation,Bib_general,Bib_Spectroscopy,Bib_interferometry,Bib_StellarEvolution,ThesisBibTexRefs}


\label{lastpage}

\end{document}